\documentclass[times]{elsarticle}
\usepackage[utf8]{inputenc}

\usepackage[margin = 1in]{geometry}
\usepackage{booktabs}
\usepackage{amsmath}
\usepackage{hyperref}
\usepackage{amssymb}
\hypersetup{colorlinks=true, linkcolor=blue, citecolor=blue, urlcolor=blue}
\usepackage[]{natbib}
\usepackage{soul}
\usepackage{multirow}
\usepackage{diagbox}
\usepackage{comment}
\usepackage{float}

\date{\today}
\journal{ }

\begin{document}

\begin{frontmatter}
    \title{Homogenizing elastic properties of large digital rock images by combining CNN with hierarchical homogenization method}

    \author[label1]{ Rasool Ahmad \corref{cor1}}
    \ead{rasool@stanford.edu}
    \author[label2]{ Mingliang Liu }
    \ead{mliu9@stanford.edu}
    \author[label3]{Michael Ortiz}
    \ead{ortiz@aero.caltech.edu}
    \author[label2]{Tapan Mukerji}
    \ead{mukerji@stanford.edu}
    \author[label1]{Wei Cai}
    \ead{ caiwei@stanford.edu }
    \address[label1]{Micro and Nano Mechanics Group, Department of Mechanical Engineering,
        Stanford University, CA 94305, USA}
    \address[label2]{Department of Energy Science and Engineering, Stanford University, CA 94305, USA}
    \address[label3]{Division of Engineering and Applied Science, California Institute of Technology, CA 91125 Pasadena, USA}
    \cortext[cor1]{corresponding author}

    \begin{abstract}
        Determining effective elastic properties of rocks from their pore-scale digital images is a key goal of digital rock physics (DRP). Direct numerical simulation (DNS) of elastic behavior, however, incurs high computational cost; and surrogate machine learning (ML) model, particularly convolutional neural network (CNN), show promises to accelerate homogenization process. 3D CNN models, however, are unable to handle large images due to memory issues. To address this challenge, we propose a novel method that combines 3D CNN with hierarchical homogenization method (HHM). The surrogate 3D CNN model homogenizes only small subimages, and a DNS is used to homogenize the intermediate image obtained by assembling small subimages. The 3D CNN model is designed to output the homogenized elastic constants within the Hashin-Shtrikman (HS) bounds of the input images. The 3D CNN model is first trained on data comprising equal proportions of five sandstone (quartz mineralogy) images, and, subsequently, fine-tuned for specific rocks using transfer learning. The proposed method is applied to homogenize the rock images of size $300\times 300 \times 300$ and $600\times 600\times 600$ voxels, and the predicted homogenized elastic moduli are shown to agree with that obtained from the brute-force DNS. The transferability of the trained 3D CNN model (using transfer learning) is further demonstrated by predicting the homogenized elastic moduli of a limestone rock with calcite mineralogy. The surrogate 3D CNN model in combination with the HHM is thus shown to be a promising tool for the homogenization of large 3D digital rock images and other random media.

    \end{abstract}

\end{frontmatter}

\section{Introduction}
\label{sec_intro}

Digital Rock Physics (DRP) aims to complement/replace the expensive laboratory experiments to characterize the relevant properties of reservoirs directly from the pore-scale digital images of the constituent rocks~\citep{Arns2001, Dvorkin2011, Andra2013, Andra2013a, Saxena2017, Saxena2019}. The DRP workflow starts from the acquisition of the three-dimensional digital images of rocks using modern imaging techniques such as micro computed-tomography (micro-CT). The digital image is then segmented into its different constituents and pores using an image processing tool. A physical simulation is then performed numerically on the segmented digital images to obtain the desired effective properties of the rock. The effective elastic properties, e.g. bulk and shear moduli, are important properties of rocks to fully characterize the reservoirs and is the focus of the present work.

Rocks are quintessential examples of random media in which microstructures (distribution of constituents materials and pores) span multiple length scale. Resolution of the finer length-scale features of the microstructure is necessary to obtain accurate results from the numerical simulations. On the other hand, a large field-of-view (FOV) is needed to sample statistically representative volume to  obtain well-converged results. Satisfying both conditions is becoming increasingly possibly with the advancement in imaging technology. Furthermore, machine learning-based fusion of high resolution and small FOV scanning electron microscopy (SEM) images with low resolution and large FOV micro-CT images yields huge images (with multibillion voxels) with high resolution and large FOV~\citep{Liu2022a}. However, performing brute-force numerical simulations on such huge images to determine effective properties remains out of reach due to requirement of prohibitively expensive requirement of computational resources $-$ in terms of both computation time and memory to store images and intermediate computation steps.

Domain decomposition-based methods are commonly utilized to efficiently carry out the numerical simulations on huge computational domains (digital images in the present context). The basic idea of domain decomposition methods is to first divide the huge image under investigation into multiple small computationally tractable subimages. The numerical simulations are performed on individual subimages to obtain their effective properties. The final step is to integrate the homogenized properties of the individual subimages to determine the effective property of the original huge image. One way to realize this last integration is by simply computing the simple arithmetic/geometric/harmonic mean of the properties of the individual subimages. Hierarchical homogenization method (HHM) is another principled renormalization method-inspired procedure to perform the last step of the  integration to find accurate effective properties of the huge domain. In HHM, the last integration step is carried out by assembling the subimages with corresponding effective properties and performing the numerical simulation on much smaller partially homogenized domain. Recently,~\citet{ahmadComputationEffectiveElastic2023} systematically analyzed the errors incurred by the HHM, and provide a procedure to obtain accurate and unbiased effective elastic moduli of digital rocks.

In domain decomposition methods such as HHM, the computational cost associated with the last integration step is negligible. The majority of the computational cost in HHM is incurred during the homogenization of the subimages by performing relatively expensive numerical simulations. Thus, replacing the expensive numerical simulations with cheaper deep learning-based surrogate models to predict the effective elastic moduli of subimages would significantly increase the overall computational efficiency of the HHM, and is the main focus of the present work.

Deep learning-based surrogate models have found wide-ranging applications in predicting the physical properties of digital rocks. The strength of deep learning methods originates from their ability to learn important features from raw data to predict the target properties. Convolutional Neural Networks (CNN) is at the core of most prominent surrogate deep leaning models for DRP applications which operate directly on the digital images to predict the desired properties. For instance, CNN-based models have been used for image segmentation~\citep{Karimpouli2019a, Phan2021, Cao2022, Niu2020}, predicting the flow properties and permeability~\citep{Wu2018b, Araya-Polo2020, Tian2020, Santos2020, Santos2021, Rizk2021,liuHierarchicalHomogenizationDeepLearningBased2023}, predicting mechanical properties~\citep{Cao2022, Cilli2022, Cui2021, Karimpouli2019, Saad2019, Rabbani2020, Eidel2023}, enhancing the resolution of digital images~\citep{Liu2022a, Wang2019a, Tawfik2022, Wang2019b}, image reconstruction~\citep{Li2018b, Mosser2018}, classification~\citep{DosAnjos2021}.

The application of CNN to large three-dimensional digital images is severely limited by the excessive memory requirements to fit the images on graphical processing units (GPUs). Large images restrict the batch size and lead to inefficient training of the model. To mitigate this memory issue, \citet{Kashefi2021} use PointNet instead of CNN to predict the permeability from rock images. PointNet operates on the point cloud representation of the interface between pore and mineral phases and thus requires much lower memory than the full image. Furthermore, \citet{Santos2021} propose a multiscale CNN to predict permeability of rock images with more than $512^3$ voxels. However, the large image size results in the smaller number of training data which can lead to overfitting and poor generalizability of the trained model.

In this work, we propose a novel hybrid approach, CNN-HHM, which combines deep learning model and direct numerical simulation (DNS) in the framework of HHM to efficiently determine the effective elastic moduli of large digital rock images. The framework starts with the partitioning of a large rock image of size $N\times N\times N$ voxels into $(N/n)^3$ disjoint smaller subimages of size $n\times n \times n$ voxels. A surrogate 3D-CNN model is trained to predict the effective isotropic elastic moduli (bulk and shear) of subimages which are numerous to train a deep learning model. To build the training data, the effective elastic moduli of the subimages are determined by solving the elasticity problem using an efficient FFT-based numerical solver. The main aspect of our 3D-CNN implementation is that it does not directly predict the bulk and shear moduli from the segmented subimages; instead the 3D-CNN model outputs two scalars $f_K$ and $f_\mu$ having values between 0 and 1. The two numbers $f_K$ and $f_\mu$ specify the effective bulk and shear moduli with respect to the respective lower and upper Hashin-Strikman bounds. This particular feature of the model results in high accuracy and excellent generalizability of the trained model. The surrogate CNN model is pretrained on a mixed data set containing equal proportion of five sandstone rock (quartz mineralogy) images of size $75\times 75 \times 75$ voxels. We, subsequently, use transfer learning to fine tune the surrogate CNN model for each rock separately.

The trained 3D-CNN model is then integrated into the HHM scheme to determine the effective elastic moduli of large rocks. In particular, the effective moduli of the subimages are obtained using the trained 3D-CNN model instead of DNS. The substitution of computationally costly numerical simulation with the cheaper surrogate 3D-CNN model leads to significant increase in the computational efficiency of the overall HHM scheme. The second and the final homogenization of much smaller intermediate assembled image is performed by direct physical simulation which has negligible computational cost. The proposed framework is shown to predict the effective elastic moduli of digital rock images of size $300\times 300\times 300$ and $600\times 600\times 600$ voxels in excellent agreement with DNS values. The CNN-HHM approach with transfer learning is then shown to predict the elastic moduli of limestone (calcite mineralogy) with reasonable accuracy. Thus, the proposed CNN-HHM approach is shown to be a promising tool for the efficient determination of the effective elastic moduli of large digital rock images with varies microstructure and mineral composition.

The rest of the manuscript is organized as follows. In Section~\ref{sec:methods}, we present the details of the hierarchical homogenization method (HHM) and the surrogate 3D-CNN model which forms the component of the CNN-HHM approach. In Section~\ref{sec:results}, we present the results of the proposed CNN-HHM approach. Section~\ref{sec:conclusion}summarizes and discusses the main findings of the present work.

\section{Methods}
\label{sec:methods}
In this section, we first briefly describe the hierarchical homogenization method (HHM) closely following~\citet{ahmadComputationEffectiveElastic2023}. We then present the details of the surrogate 3D-CNN model including the architecture, the loss function used to train the model, and the generation of the training, validation and testing data.

\subsection{Hierarchical Homogenization Method (HHM)}
\label{sec:HHM}
Hierarchical homogenization method (HHM) is a renormalization inspired approximate method to determine the homogenized properties of large computational domains. This method is especially helpful for digital rock physics application, as requirement of high resolution and large field of view result in large images that cannot be handled by brute force direct numerical simulations. Instead of solving the elasticity PDEs in the large image, HHM adopts the renormalization inspired approach where the large computational domain is scaled down by successive coarse-graining.

\begin{figure}[ht!]
    \centering
    \includegraphics[width=\linewidth]{"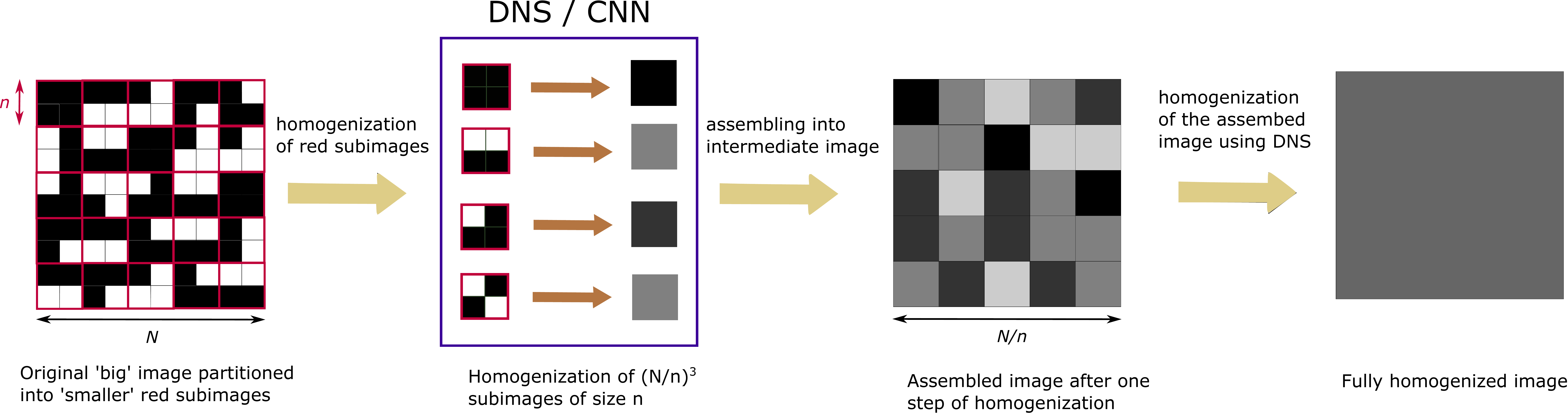"}
    \caption{Schematic of the HHM scheme. The large image is partitioned into smaller red subimages. The red subimages are homogenized separately by direct numerical simulation (DNS) or a surrogate convolutional network (CNN). The homogenized subimages are assembled to obtain a partially homogenized image. The assembled partially homogenized image is again homogenized by directly solving the elasticity PDEs to find the final homogenized elastic constants.}
    \label{fig:HHM}
\end{figure}

HHM starts with the partitioning of a 3D large image of size (number of voxels) $N \times N \times N$ into $(N/n)^3$ smaller subimages of size $n \times n \times n$ as schematically depicted in red in the first panel of Figure~\ref{fig:HHM}. Each voxel of the image represents one of the two phases (pore and mineral in our case). The two phases are considered to be homogeneous isotropic elastic materials with their own elastic constants. We then homogenize the red subimages and replace them with one voxel of corresponding homogenized elastic constants. The subimages are homogenized by solving the elasticity PDEs using periodic boundary conditions. In the next step, we assemble the homogenized subimges to obtain a partially homogenized version of the original large image, as shown in the third panel of Figure~\ref{fig:HHM}. The partially homogenized image is much smaller ($(N/n)^3$ voxels) in size compared to original large image, and is amenable to brute direct numerical simulations. Furthermore, the partially homogenized image contains the low-frequency information of the original image and the high-frequency local information are summarized into the homogenized properties of subimages. The assembled image is again homogenized by solving the elasticity PDEs using periodic boundary conditions to find the final homogenized isotropic elastic constants.

\begin{figure}[ht!]
    \centering
    \includegraphics[width=\linewidth]{"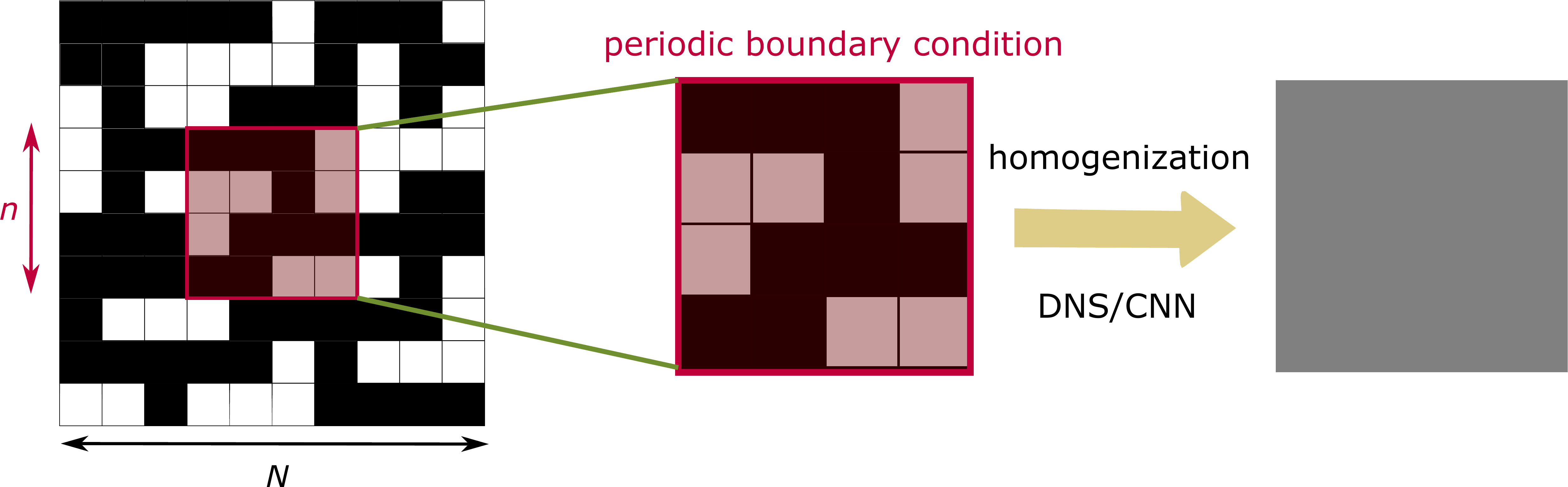"}
    \caption{Schematic of the homogenization of a single subimage. The subimage is composed of two phases (pore and mineral). The segmented subimage is homogenized either by solving the elasticity PDEs (DNS) under periodic boundary conditions or using a surrogate CNN model.}
    \label{fig:subimage_homogenization}
\end{figure}

In the original HHM scheme, the individual subimages are extracted from the large image and homogenized by direct numerical simulation (DNS) as depicted in Figure~\ref{fig:subimage_homogenization}. The subimages are subjected to a periodic boundary condition and the equilibrium stress and strain field is determined by solving the elasticity PDEs using FFT-based elasticity solver. The effective elastic stiffness matrix is then determined by relating the average stress and strain in the subimages. To find all the components of the effective stiffness matrix, the elasticity PDEs are solved for six independent boundary conditions. For more details, reader are directed to~\citet{ahmadComputationEffectiveElastic2023}.

Thus, HHM enables the homogenization of a large image without solving the elasticity equations in the large image itself. Since most of the time of the HHM is spent to homogenize $(N/n)^3$ subimages, a fast surrogate CNN model to obtain the homogenized elastic constant of smaller subimages will significantly accelerate the whole HHM scheme.

\subsection{Details of the surrogate three-dimensional CNN model}
\label{sec:3D-CNN}
As depicted in Figure~\ref{fig:subimage_homogenization}, the smaller red subimages are extracted from the large image and processed by a surrogate 3D CNN model. In this work, our 3D CNN model is designed to take images of size $75 \times 75 \times 75$ as input and output the effective bulk and shear moduli of the subimage. The input images are already segmented into two phases of pore and mineral.

\begin{figure}[ht!]
    \centering
    \includegraphics[width=\linewidth]{"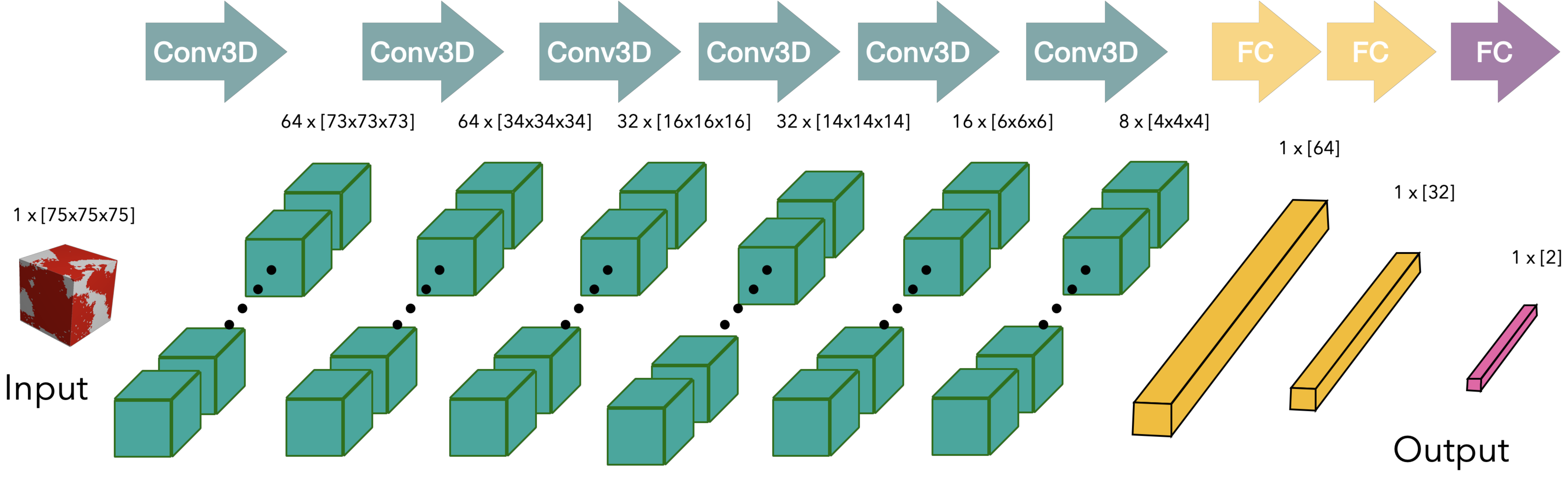"}
    \caption{Schematic of the 3D CNN model. The input to the model is a segmented (pore and mineral) 3D image of size $75 \times 75 \times 75$). The model is composed mainly of 3D convolution and fully connected linear layers with additional batchnorm, average pooling, non-linear activation layers. The output of the model is an array of two numbers $f_K$ and $f_\mu$ which lie between 0 and 1. The details of the CNN and fully connected layers are presented in Table~\ref{tab:cnn_model}}
    \label{fig:cnn_model}
\end{figure}

The architecture of the surrogate CNN model is depicted schematically in Figure~\ref{fig:cnn_model}; and the details of the various layers involved in the model are presented in Table~\ref{tab:cnn_model}. The input image is first passed through a series of 3D convolution layers with batch normalization, average pooling and PRelu non-linear activation layers. The output of the last 3D convolution layer is then flattened and passed through a series of fully connected linear layers with batch normalization and PRelu non-linear activation layers. The output of the final linear layer is passed through $0.5 ( \tanh(\cdot) + 1)$ function to constrain the values between 0 and 1. Thus, the final output of the surrogate CNN model is an array of two numbers $f_K\in (0,1)$ and $f_\mu \in (0,1)$ which indicate the effective moduli location between their respective Hashin-Shtrikman bounds. Specifically, the effective bulk and shear moduli are obtained by using the following equations
\begin{equation}
    \begin{aligned}
        K_\text{eff}   & = f_K K^\text{HS}_\text{lower} + (1-f_K) K^\text{HS}_\text{upper},         \\
        \mu_\text{eff} & = f_\mu \mu^\text{HS}_\text{lower} + (1-f_\mu) \mu^\text{HS}_\text{upper},
    \end{aligned}
    \label{eq:effective_moduli}
\end{equation}
where $K^\text{HS}_\text{lower}$ and $K^\text{HS}_\text{upper}$ are the lower and upper Hashin-Shtikman bounds of the bulk modulus, and $\mu^\text{HS}_\text{lower}$ and $\mu^\text{HS}_\text{upper}$ are the lower and upper Hashin-Shtikman bounds of the shear modulus. The use of the Hashin-Shtrikman bounds to inform the surrogate CNN model improves the accuracy of the prediction of the effective elastic moduli, and enables the transferability of the model to different rock types as will be demonstrated later in Section~[]. Furthermore, the Hashin-Shtrikman bounds are cheap themselves to compute from the information of the volume fraction of the two phases (porosity in the case of rock) in the subimages and their respective bulk and shear moduli~\citep{Hashin1963}, as
\begin{equation}
    \begin{aligned}
        K^\text{HS}_\text{upper}   & = K_\text{mineral} + \frac{1-\phi}{\left(K_\text{pore} - K_\text{mineral}\right)^{-1} + \phi \left(K_\text{mineral} + \frac{4}{3} \mu_\text{mineral} \right)},                                                                             \\
        K^\text{HS}_\text{lower}   & = K_\text{pore} + \frac{\phi}{\left(K_\text{mineral} - K_\text{pore}\right)^{-1} + \left(1-\phi\right) \left(K_\text{pore} + \frac{4}{3} \mu_\text{pore} \right)},                                                                         \\
        \mu^\text{HS}_\text{upper} & = \mu_\text{mineral} + \frac{1-\phi}{\left(\mu_\text{pore} - \mu_\text{mineral}\right)^{-1} + \frac{2\phi (K_\text{mineral} + 2\mu_\text{mineral})}{5\mu_\text{mineral}\left( K_\text{mineral} + \frac{4}{3}\mu_\text{mineral}\right) } }, \\
        \mu^\text{HS}_\text{lower} & = \mu_\text{pore} + \frac{\phi}{\left(\mu_\text{mineral} - \mu_\text{pore}\right)^{-1} + \frac{2(1-\phi) (K_\text{pore} + 2\mu_\text{pore})}{5\mu_\text{pore}\left( K_\text{pore} + \frac{4}{3}\mu_\text{pore}\right) } },
    \end{aligned}
\end{equation}
where $\phi$ is the porosity, $K_\text{pore}$ and $\mu_\text{pore}$ are the bulk and shear moduli of the pore phase, and $K_\text{mineral}$ and $\mu_\text{mineral}$ are the bulk and shear moduli of the mineral phase.

\begin{table}[ht!]
    \centering
    \caption{The architecture of the surrogate CNN model. The model comprises two parts: the convolution part and the fully connected part as shown in Figure~\ref{fig:cnn_model} }
    \label{tab:cnn_model}
    \vspace{2mm}
    The convolution part
    \begin{tabular}{c c c c c}
        \toprule
        Layer                 & Input size                          & Kernel size & Number of filters & Output size                         \\
        \midrule
        Image                 & $ 1 \times 75 \times 75 \times 75$  & -           & -                 & -                                   \\
        Batchnorm3D           & $ 1 \times 75 \times 75 \times 75$  & -           & -                 & $ 1 \times 75 \times 75 \times 75$  \\
        Conv3D + PRelu        & $ 1 \times 75 \times 75 \times 75$  & 3           & 64                & $ 64 \times 73 \times 73 \times 73$ \\
        Dropout3D(prob = 0.5) & $ 64 \times 73 \times 73 \times 73$ & -           & -                 & $ 64 \times 73 \times 73 \times 73$ \\
        Batchnorm3D           & $ 64 \times 73 \times 73 \times 73$ & -           & -                 & $ 64 \times 73 \times 73 \times 73$ \\
        Conv3D + PRelu        & $ 64 \times 73 \times 73 \times 73$ & 6           & 64                & $ 64 \times 68 \times 68 \times 68$ \\
        AvgPool3D             & $ 64 \times 68 \times 68 \times 68$ & 2           & -                 & $ 64 \times 34 \times 34 \times 34$ \\
        Batchnorm3D           & $ 64 \times 34 \times 34 \times 34$ & -           & -                 & $ 64 \times 34 \times 34 \times 34$ \\
        Conv3D + PRelu        & $ 64 \times 34 \times 34 \times 34$ & 3           & 32                & $ 32 \times 32 \times 32 \times 32$ \\
        AvgPool3D             & $ 32 \times 32 \times 32 \times 32$ & 2           & -                 & $ 32 \times 16 \times 16 \times 16$ \\
        Batchnorm3D           & $ 32 \times 16 \times 16 \times 16$ & -           & -                 & $ 32 \times 16 \times 16 \times 16$ \\
        Conv3D + PRelu        & $ 32 \times 16 \times 16 \times 16$ & 3           & 32                & $ 32 \times 14 \times 14 \times 14$ \\
        Dropout3D(prob = 0.5) & $ 32 \times 14 \times 14 \times 14$ & -           & -                 & $ 32 \times 14 \times 14 \times 14$ \\
        Batchnorm3D           & $ 16 \times 14 \times 14 \times 14$ & -           & -                 & $ 32 \times 14 \times 14 \times 14$ \\
        Conv3D + PRelu        & $ 16 \times 14 \times 14 \times 14$ & 3           & 16                & $ 16 \times 12 \times 12 \times 12$ \\
        AvgPool3D             & $ 16 \times 12 \times 12 \times 12$ & 2           & -                 & $ 16 \times 6 \times 6 \times 6$    \\
        Batchnorm3D           & $ 16 \times 6 \times 6 \times 6$    & -           & -                 & $ 16 \times 6 \times 6 \times 6$    \\
        Conv3D + PRelu        & $ 16 \times 6 \times 6 \times 6$    & 3           & 8                 & $ 8 \times 4 \times 4 \times 4$     \\
        Dropout3D(prob = 0.5) & $ 8 \times 4 \times 4 \times 4$     & -           & -                 & $ 8 \times 4 \times 4 \times 4$     \\
        \bottomrule
    \end{tabular}

    \vspace{2mm}
    The fully connected part

    \begin{tabular}{c c c}
        \toprule
        Layer                   & Input size     & Output size     \\
        \midrule
        Batchnorm1D             & $1 \times 512$ & $ 1 \times 512$ \\
        Fully connected + PRelu & $1 \times 512$ & $ 1 \times 64$  \\
        Dropout1D(prob = 0.5)   & $1 \times 64$  & $ 1 \times 64$  \\
        Batchnorm1D             & $1 \times 64$  & $ 1 \times 64$  \\
        Fully connected + PRelu & $1 \times 64$  & $ 1 \times 32$  \\
        Batchnorm1D             & $1 \times 32$  & $ 1 \times 32$  \\
        Fully connected         & $1 \times 32$  & $ 1 \times 2$   \\
        0.5(Tanh + 1)           & $1 \times 2$   & $ 1 \times 2$   \\
        \bottomrule
    \end{tabular}
\end{table}

\subsection{Details of the training of the surrogate CNN model}
\label{sec:details_training}
\begin{figure}
    \centering
    \includegraphics[width=0.5\linewidth]{"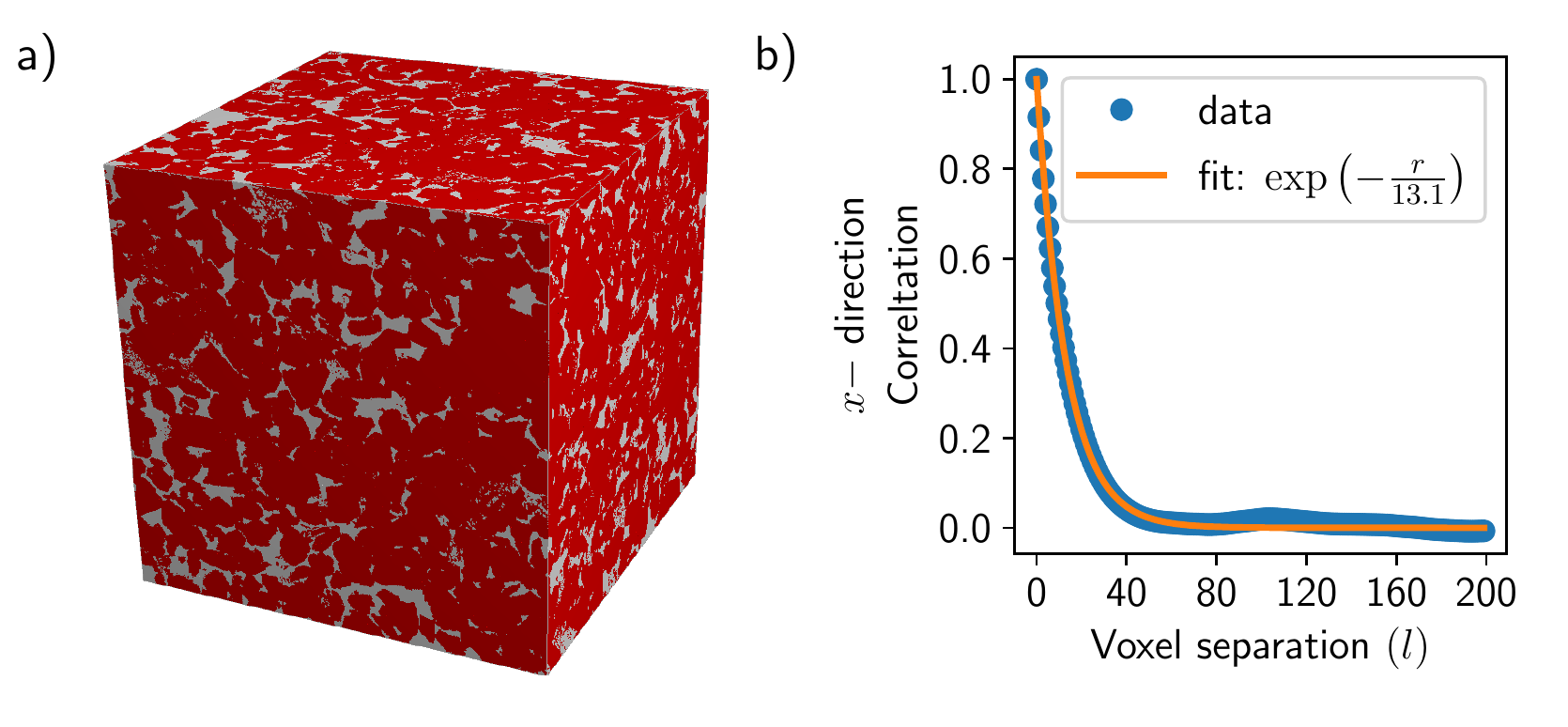"}
    \caption{(a) Segmented image of rock sample B1 of size $900 \times 900 \times 900$ voxels. Red color represents quartz mineral and white color denotes pore. (b) Two point correlation function of the segmented image of rock sample B1 along the $x-$direction. Blue dots are computed using the segmented image and the orange curve is the best fit to exponential function $\exp\left( - l/\xi \right)$, with $\xi$ being the correlation length. $\xi$ is expressed in the units of numbers of voxel which is approximately 2 $\mu$m in length.}
    \label{fig:image_correlation_B1}
\end{figure}

The surrogate CNN model is trained on a mixed dataset comprising equal proportion of five sandstone rock images of size $75\times 75 \times 75$ voxels. The five sandstone used in this work are the same as those used in~\citet{Saxena2019} and~\citet{ahmadComputationEffectiveElastic2023}: two (B1 and B2) are from the Berea Formation, subangular to subrounded Mississippian-age sandstone; two (FB1 and FB2) from the Fontainebleau Formation, subrounded to rounded Oligocene age sandstone; and one (CG) from the Castlegate Formation, subangular to subrounded Mesozoic sandstone. All the rocks were imaged with a micro-CT scanner at the image resolution of approximately 2 $\mu$m (the physical size of each voxel). The X-ray diffraction (XRD) analysis demonstrates that all samples mainly consist of quartz mineral along with trace amounts of feldspar, calcite and clay. For simplicity, we assume single mineralogy where all minerals were treated as quartz. The rock images are segmented into two phases (pore and mineral) using the Otsu thresholding method~\citep{otsuThresholdSelectionMethod1979}. Quartz phase is assigned bulk modulus of 36 GPa and shear modulus of 45 GPa; and pore phase's bulk and shear moduli are assumed to be 0 GPa ~\citep{mavko2003rock} (for numerical stability, pore's bulk and shear moduli are taken to be $10^{-5}$ GPa). Figure~\ref{fig:image_correlation_B1} shows the segmented image of rock sample B1 and the two-point correlation function along the $x-$direction. The porosity and correlation function of all five sandstone rocks are presented in Table~\ref{tab:rock_samples}. The rock porosity ranges from 3.45\% to 22.20\% and all five rocks have similar correlation length in all three direction indicating their isotropic structures.

\begin{table}[ht!]
    \centering
    \caption{Statistical properties of the rock samples, including porosity and correlation lengths $\xi$ in $x$, $y$, and $z$ directions. }
    \begin{tabular}{ c  c c c c }
        \toprule
        Rock sample & porosity & $\xi_x$ & $\xi_y$ & $\xi_z$ \\
        \midrule
        B1          & 16.51\%  & 13.1    & 12.8    & 13.4    \\
        B2          & 19.63\%  & 14.2    & 15.3    & 14.5    \\
        CG          & 22.20\%  & 10.7    & 8.5     & 11.8    \\
        FB1         & 9.25\%   & 14.4    & 16.1    & 15.6    \\
        FB2         & 3.45\%   & 15.9    & 15.1    & 15.1    \\
        \bottomrule
    \end{tabular}
    \label{tab:rock_samples}
\end{table}

\begin{figure}[ht!]
    \centering
    \includegraphics[width=0.6\textwidth]{"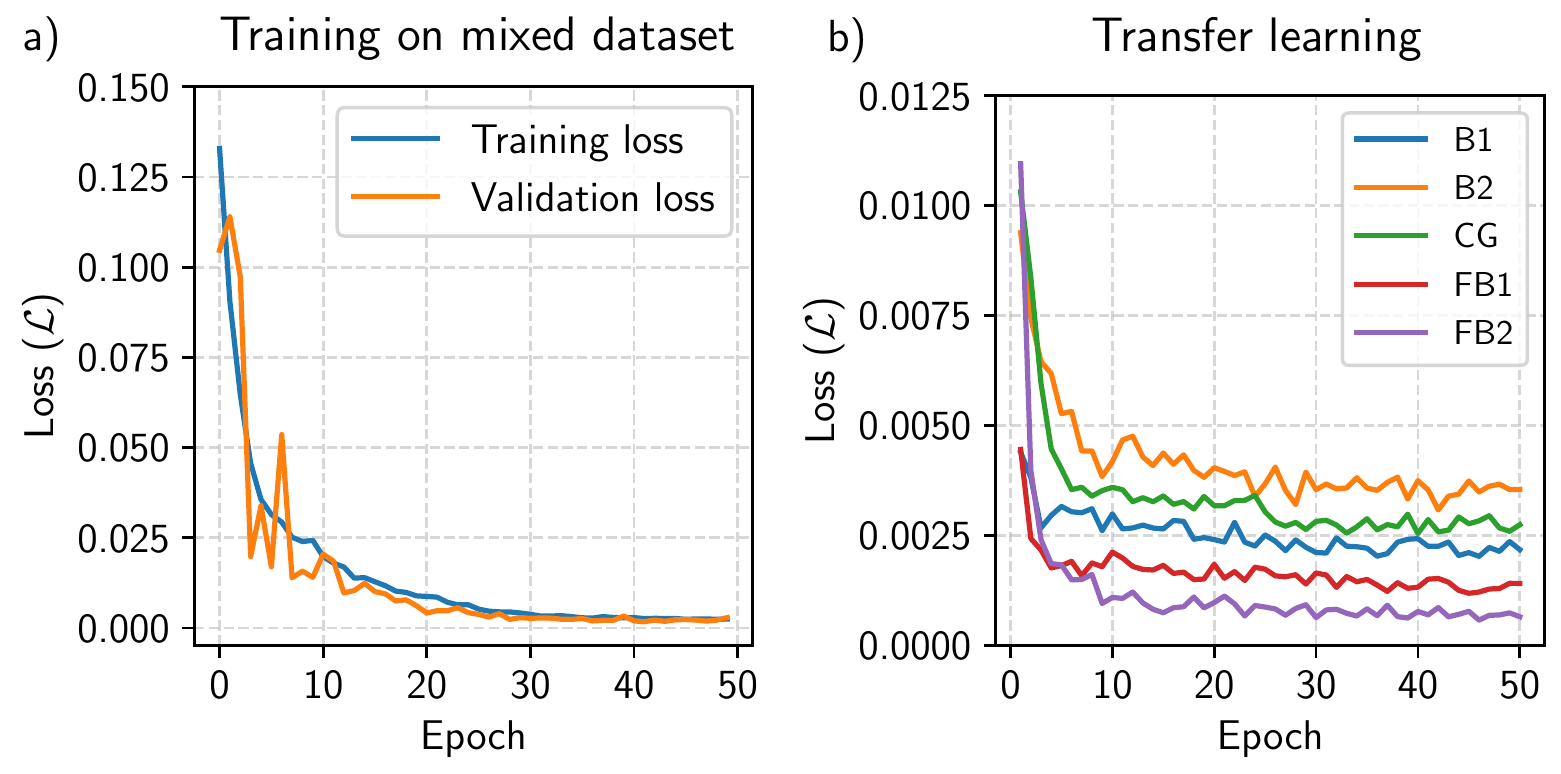"}
    \caption{ Variation of the losses $(\mathcal{L})$ during (a) training on mixed dataset and (b) transfer learning. In (a), both training and validation loss functions are shown. In (b), only the training loss functions are shown for five different sandstones.}
    \label{fig:loss_function}
\end{figure}

The training data for the surrogate CNN model is generated by homogenizing the subimages by directly solving the elasticity PDEs using an FFT-based elasticity solver as implemented in GeoDict~\citep{Elastodict, Kabel2013, Kabel2016, Schneider2016}. The size of the training data is 2800 which contains all five sandstone samples in equal proportion. Moreover, we use 1000 random rock images from mixed dataset for the validation of the surrogate model. The surrogate CNN model is trained in supervise manner using the following mean squared loss function
\begin{align}
    \mathcal{L} = \frac{1}{N_\text{batch}} \sum_{i=1}^{N} \left( \left| \left| K_\text{eff, i}^\text{DNS} - K_\text{eff, i}^{\text{pred}} \right| \right|_2^2  +  \left| \left| \mu_\text{eff, i} ^\text{DNS} - \mu_\text{eff, i}^{\text{pred}} \right| \right|_2^2 \right),
\end{align}
where $K_\text{eff}^{DNS}$ and $\mu_\text{eff}^{DNS}$ are the effective bulk and shear moduli obtained from DNS, $K_\text{eff}^{\text{pred}}$ and $\mu_\text{eff}^{\text{pred}}$ are the effective bulk and shear moduli predicted by the surrogate CNN model, and $N_\text{batch}$ is the batch size. The surrogate CNN model is trained using the Adam optimizer~\citep{kingmaAdamMethodStochastic2017} for 50 epochs where a learning rate of $10^{-3}$ is used for the first 25 epochs and $10^{-4}$ for the last 25 epochs. The surrogate CNN is implemented using the PyTorch deep learning framework~\citep{NIPS2019_9015} and trained on GPU. The variation of the losses during training is shown in Fig.~\ref{fig:loss_function}(a). The training loss is shown in blue and the validation loss is shown in orange. Both the training and validation losses decrease during training indicating that the model is learning to predict effective elastic moduli using images. We choose the surrogate CNN model for which the validation loss s the minimum.

After training the surrogate model on the mixed dataset, we make use of \textit{transfer learning} to fine tune the model for each rock separately. During transfer learning, we fine tune only the fully connected part of the surrogate CNN model using randomly sampled 1000 images of the rock. The fine-tuning is performed for 50 epochs with a learning rate of $10^{-3}$ for the first 25 epochs and $10^{-4}$ for the next 25 epochs. The variation of the training loss functions during transfer learning are shown in Figure~\ref{fig:loss_function}(b) for all five rocks. For all five rocks, both the training and validation loss functions decrease during the transfer learning phase. Thus, after the transfer learning step, we obtain five different surrogate CNN models, one for each sandstone.

Furthermore, the results obtained from the pretrained CNN model is presented in \ref{app:pretrained_model}.

\section{Results}
\label{sec:results}
In this section, we present the results obtained from the application of the CNN-HHM approach(surrogate CNN model combined with the HHM) to determine effective elastic moduli of sandstone images of size $300 \times 300 \times 300$ and $600 \times 600 \times 600$ voxels. We also apply the developed framework to limestone with carbonate minerals. All the results obtained from the CNN-HHM approach are compared with the results obtained from the DNS.

\subsection{Homogenization of subimages of size $75 \times 75 \times 75$ voxels using the surrogate CNN model}
\begin{figure}
    \centering
    \includegraphics[width=\textwidth]{"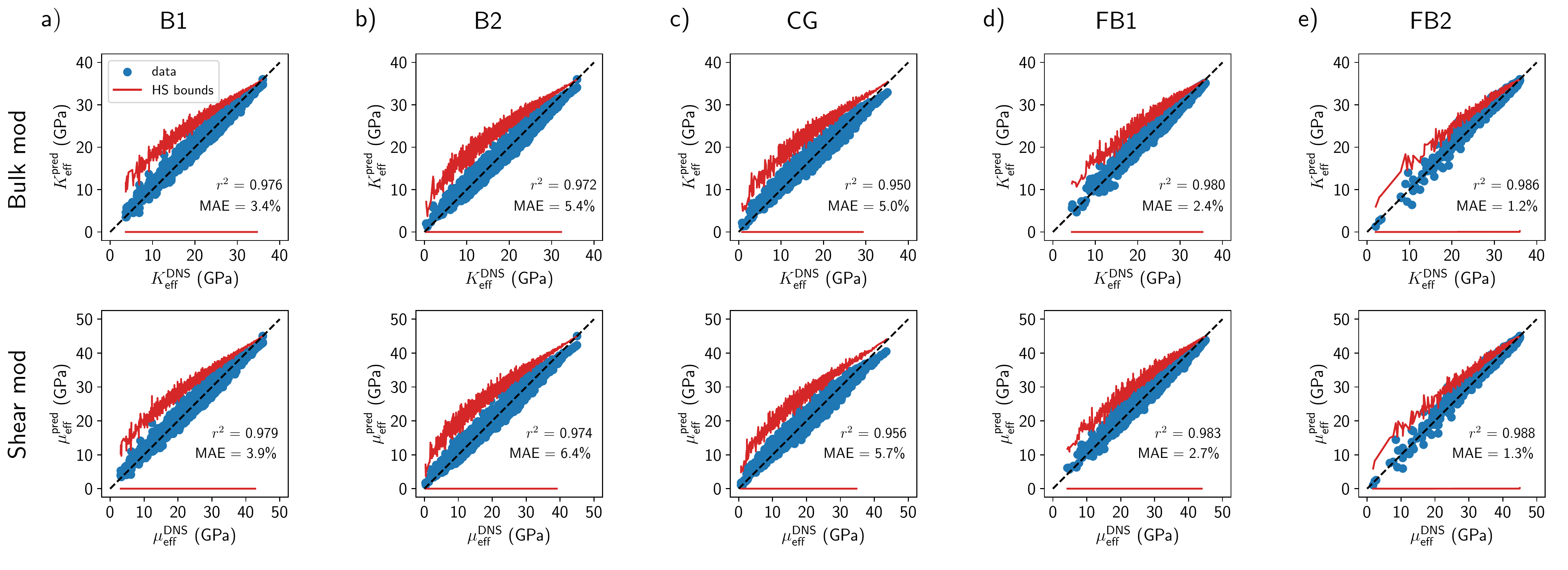"}
    \caption{Parity plot of the effective elastic moduli predicted from the fine-tuned surrogate CNN model vs that obtained from DNS for the subimages of size $75 \times 75 \times 75$ voxels. The results are shown for five different sandstones: (a) B1, (b) B2, (c) CG, (d) FB1, and (e) FB2. The blue dots represent the elastic moduli of the individual rocks; red lines show the HS bounds of the individual rocks; and the black dashed line is $45^\circ$ line. Each figure, additionally, contains $r^2$ value and mean absolute error (MAE) of the prediction.}
    \label{fig:test_result_fine_tune_True}
\end{figure}

We first show the capability of the trained CNN model to predict the homogenized elastic moduli of the small subimages of size $75 \times 75 \times 75$ voxels on which the model was trained. We use rock-specific CNN models obtained from transfer learning to predict effective elastic moduli of 4096 images of each sandstone. Figure~\ref{fig:test_result_fine_tune_True} shows the comparison of the effective elastic moduli predicted from the surrogate CNN model and that obtained from DNS. The blue dots denote the effective elastic moduli of the individual rock image and red line depicts the corresponding Hashin-Shtrikman bounds. The predicted effective elastic moduli from the surrogate CNN model are, thus,  well within the Hashin-Shtrikman bounds.

The Figure~\ref{fig:test_result_fine_tune_True} clearly shows that the surrogate CNN model predicts the effective bulk and shear moduli of the subimages in good agreement with the DNS results. This is reflected from the fact that the blue dots are scattered about $45^\circ$ line, shown in dashed black line. The $r^2$ values and mean absolute error (MAE) of the prediction are also shown in the figure. Among all cases, the $r^2$ values remain greater than 0.956 and MAE  lower than 6.4\% for all the five rocks, further indicating the excellent predictive power of the surrogate CNN model.

\subsection{Homogenization of image sizes $300 \times 300 \times 300$ and $600 \times 600 \times 600$ voxels using CNN-HHM approach}

We now apply the CNN-HHM approach, as illustrated in Figure~\ref{fig:HHM}, to the images of size $300 \times 300 \times 300~(N=300)$ and $600\times 600\times 600~(N=600)$ voxels. The surrogate CNN model is able to predict the effective elastic moduli of the subimages of fixed size $75 \times 75 \times 75$ voxels. Therefore, we first divide the images of size $N=300$ v into 64 and images of size $ N= 600$ into 128 subimages of size $75 \times 75 \times 75$ voxels. We then determine the effective elastic moduli of the subimages using the surrogate CNN model. The subimages are then assembled to obtain a partially homogenized image of size $4\times 4 \times 4$ voxel for image size $N=300$ and of size $8\times 8\times 8$ voxels.  We again homogenize the partially homogenized image of size $4\times 4 \times 4$ voxel using DNS which is an FFT-based elasticity solver~\citep{Moulinec1994,DeGeus2017a}. To obtain 'True' effective elastic, we replace the surrogate CNN model with DNS to homogenize subimages of size $75 \times 75 \times 75$ voxels.

\begin{figure}[ht!]
    \centering
    \includegraphics[width=\textwidth]{"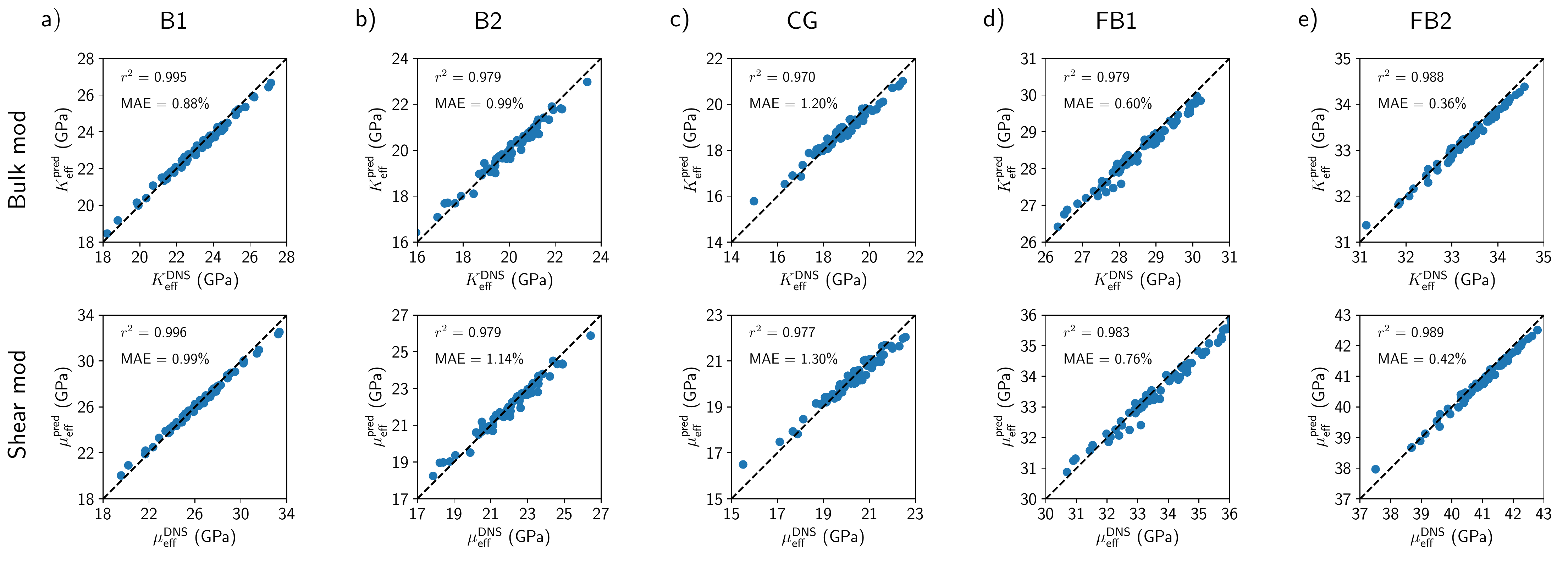"}
    \caption{Parity plot of the effective elastic moduli predicted from the CNN-HHM approach vs that obtained from DNS for the images of size $300 \times 300 \times 300$ voxels. The results are shown for five different sandstones: (a) B1, (b) B2, (c) CG, (d) FB1, and (e) FB2. The blue dots represent the elastic moduli of the individual rocks and the black dashed line is $45^\circ$ line. Each figure, additionally, contains $r^2$ value and mean absolute error (MAE) of the prediction.}
    \label{fig:HS_HHM_300}
\end{figure}

We apply the CNN-HHM approach to homogenize 64 different images of size $N=300$ voxels. Figure~\ref{fig:HS_HHM_300} shows parity plots of effective elastic moduli for all five sandstones. The blue dots representing the effective elastic moduli of the individual rock images are scattered about the $45^\circ$ line. The $r^2$ values and MAE of the prediction are also shown in the figure. Among all cases, the $r^2$ values remain greater than 0.970 and MAE lower than 1.3\% for all the five rocks. Thus, the results indicate the excellent predictive power and accuracy of the CNN-HHM.

\begin{figure}[ht!]
    \centering
    \includegraphics[width=\textwidth]{"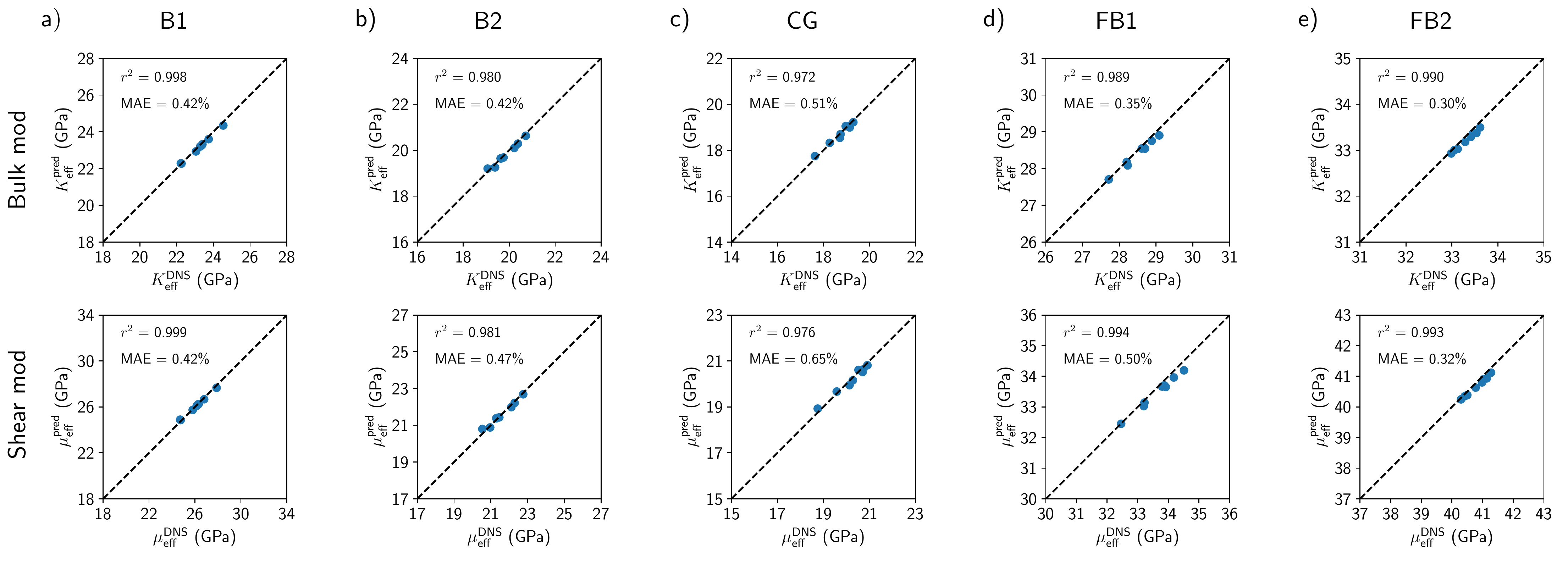"}
    \caption{Parity plot of the effective elastic moduli predicted from the CNN-HHM approach vs that obtained from DNS for the images of size $600 \times 600 \times 600$ voxels. The results are shown for five different sandstones: (a) B1, (b) B2, (c) CG, (d) FB1, and (e) FB2. The blue dots represent the elastic moduli of the individual rocks and the black dashed line is $45^\circ$ line. Each figure, additionally, contains $r^2$ value and mean absolute error (MAE) of the prediction.}
    \label{fig:HS_HHM_600}
\end{figure}

We next apply the CNN-HHM approach to predict effective bulk and shear moduli of 8 different images of size $N=600$ voxels. Figure~\ref{fig:HS_HHM_600} shows parity plots of effective elastic moduli for all five sandstones. The blue dots representing the effective elastic moduli of the individual rock images are scattered about the $45^\circ$ line. The $r^2$ values and MAE of the prediction are also shown in the figure. Among all cases, the $r^2$ values remain greater than 0.972 and MAE lower than 0.65\% for all the five rocks. We further note that the CNN prediction for rocks of size $N=75$ shows a scatter about the $45^\circ$ line, whereas the CNN-HHM predictions for rocks of size $N=300$ and $N=600$ are much tighter and show relatively small scatter about the $45^\circ$ line. This is due to the fact that the elastic moduli of larger rocks of sizes 300 and 600 are computed using multiple subimages of size 75. This averaging (in some sense) process essentially attenuates the scatter and make results much tighter.

The results presented in this section, thus, demonstrate that CNN model in combination with the HHM can be used to efficiently predict the effective elastic moduli of the large rock images.

\subsection{Homogenization of limestone rock images of size $300\times 300\times 300$ voxels using CNN-HHM approach}

\begin{figure}[ht!]
    \centering
    \includegraphics[width=0.5\textwidth]{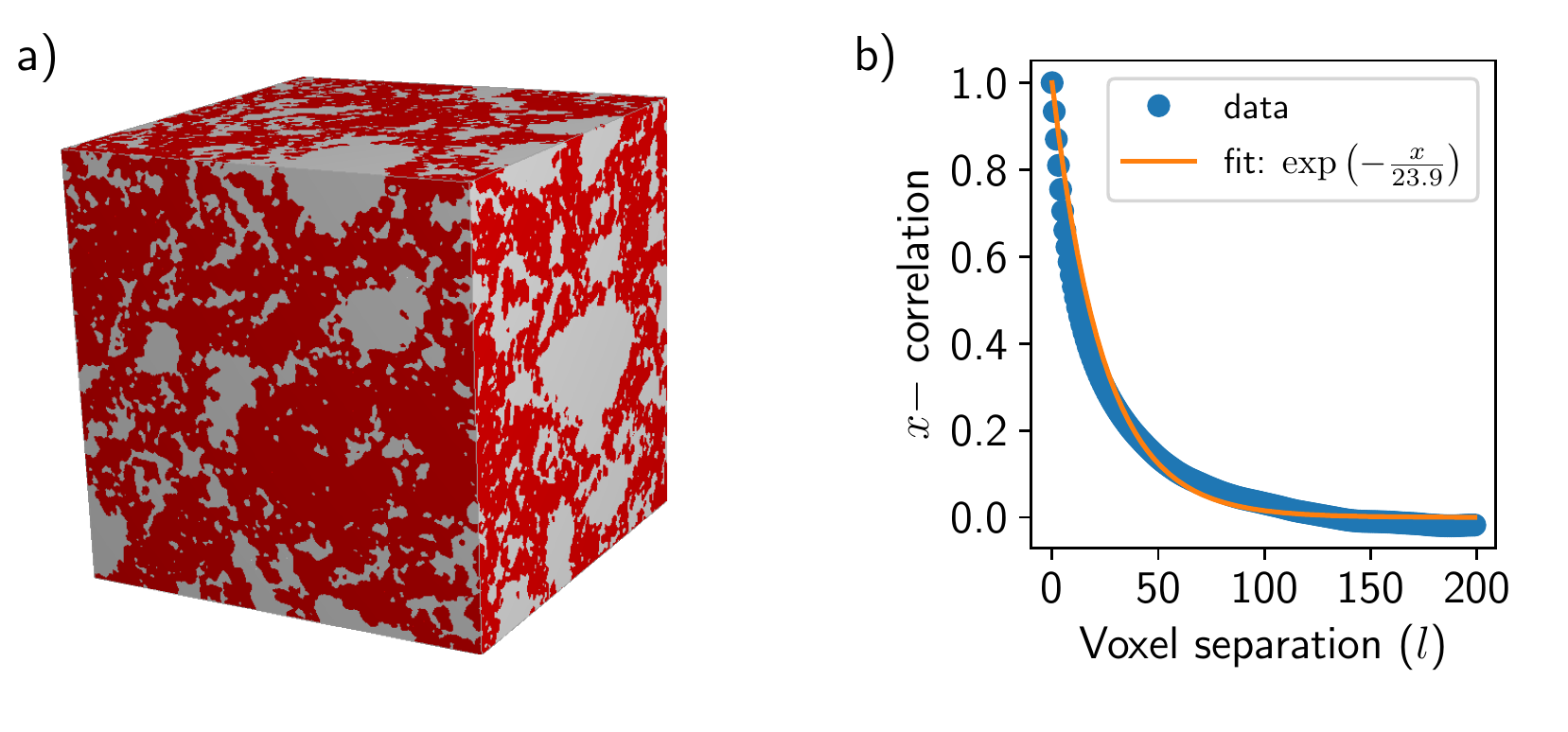}
    \caption{(a) Segmented image of limestone rock. Red color represents the calcite phase and white color represents the pores. (b) Two-point correlation function of the segmented image along the $x-$direction. Blue dots are the computed values and the orange curve the best fit to the data of the exponential function $\exp\left(-l / \xi\right)$, with $\xi$ being the correlation length. $\xi$ is expressed in the units of numbers of voxels which is approximately 2 $\mu$m in length.}
    \label{fig:limestone_structure}
\end{figure}

This section presents the application of the CNN-HHM approach to homogenize limestone images of size $300\times 300\times 300~(N=300)$ voxels. Limestones have calcite based mineralogical compositions (compared to the quartz mineralogy of the sandstone). The carbonate mineralogy is characterized by a high porosity and low elastic moduli. Figure~\ref{fig:limestone_structure}(a) shows the microstructure of the limestone used in this work. The porosity of this limestone rock is 42.38\%, much higher than the porosity of the sandstones (Table~\ref{tab:rock_samples}). Moreover, as shown in Figure~\ref{fig:limestone_structure}(b), the correlation length of the limestone rock is 23.9 voxels which is almost twice as long as that of sandstone rocks (Table~\ref{tab:rock_samples}). Additionally, the bulk and shear moduli of the calcite mineral, respectively, are 77 GPa and 32 GPa. Thus, the structural and mechanical properties of the limestone rock are significantly different from those of the sandstone rocks comprising the training dataset to pretrain the model.

\begin{figure}[ht!]
    \centering
    \includegraphics[width=0.6\textwidth]{"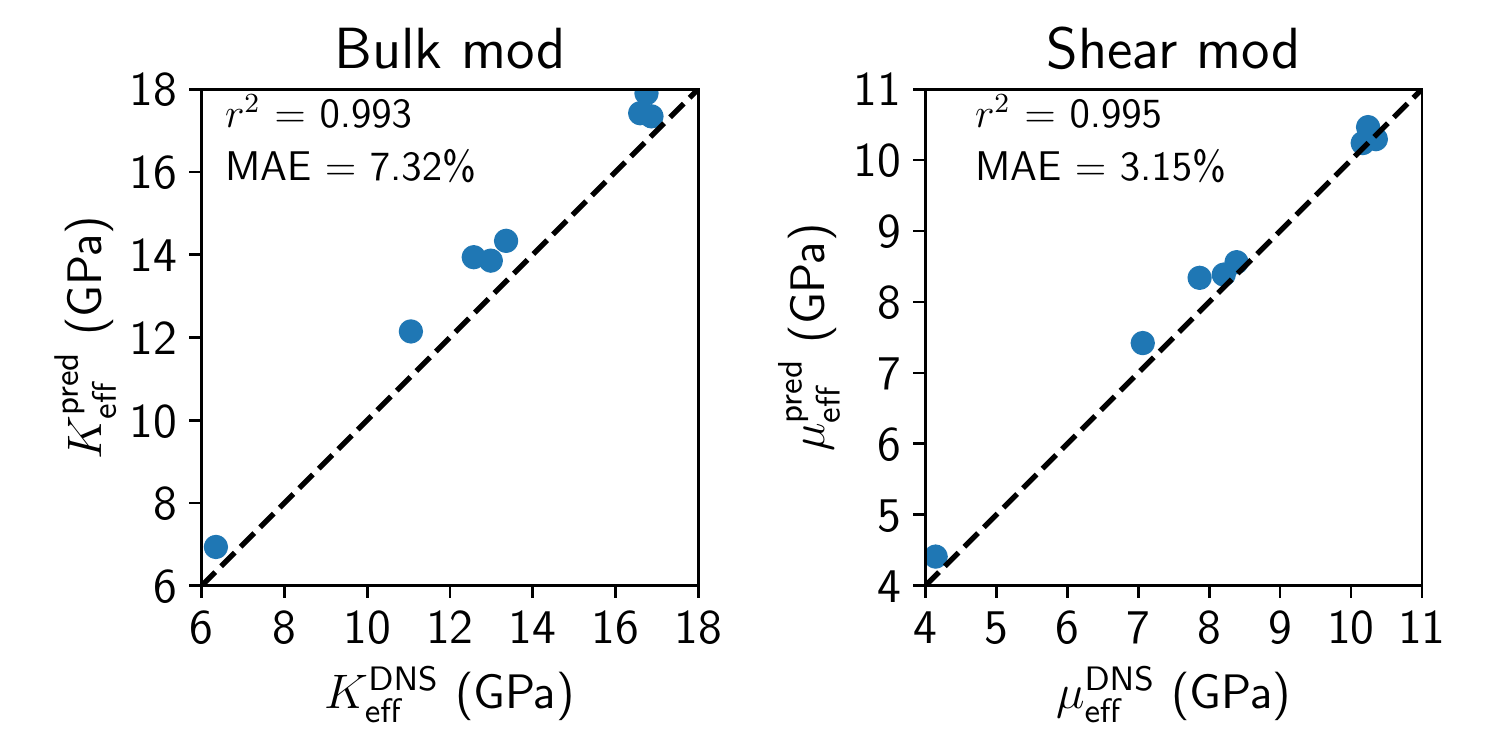"}
    \caption{Parity plot of the effective bulk and shear moduli     predicted from the CNN-HHM approach vs that obtained from DNS for the 8 limestone rock images of size $300 \times 300 \times 300$ voxels. The blue dots represent the elastic moduli of the individual rocks and the black dashed line is $45^\circ$ line. Each figure, additionally, contains $r^2$ value and mean absolute error (MAE) of the prediction.}
    \label{fig:HS_HHM_limestone}
\end{figure}

To apply the CNN-HHM approach framework to the limestone, we first use transfer learning to fine-tune the surrogate CNN model for limestone rocks that was pretrained to predict effective elastic moduli of sandstone rocks. During transfer learning, only fully-connected part of the surrogate CNN model (see Figure~\ref{fig:cnn_model}) was optimized for 50 epochs using 1000 images of limestone of size $75\times 75\times 75$ voxels. Subsequently, the surrogate model was incorporated into CNN-HHM approach framework to homogenize 8 different images of size $N=300$ voxels. Figure~\ref{fig:HS_HHM_limestone} shows parity plots of effective elastic moduli for all five sandstones. The blue dots representing the effective elastic moduli of the individual rock images are scattered about the $45^\circ$ line. The $r^2$ values and MAE of the prediction are also shown in the figure. For bulk moduli, the $r^2$ value is 0.993 and MAE is 7.3\%; for shear moduli, the $r^2$ value is 0.995 and MAE is 3.15\%. The MAE for the predicted limestone bulk moduli is somewhat larger than that for sandstone moduli. Overall, the results show that the CNN-HHM approach framework in combination with the transfer learning of the surrogate CNN model can be used to efficiently estimate the effective elastic moduli of the large rock images across compositionality.

\section{Conclusion and discussion}
\label{sec:conclusion}
In this work, we propose a hybrid CNN-HHM approach, a combination of 3D CNN and DNS, which enables and efficient estimation of the effective elastic moduli of large segmented 3D rock images. The proposed approach leverages the HHM to address the memory issue that renders the training of the 3D CNN for large 3D images infeasible. The CNN-HHM approach consists of two parts: 1) a large rock image of size $N\times N\times N$ voxels is divided into smaller disjoint subimages of size $n\times n\times n$; The subimages are then homogenized using a surrogate CNN model. 2) The homogenized subimages are assembled into smaller coarse grained image of size $N/n \times N/n \times N/n$ voxels, and subsequently homogenized using DNS to find the final effective elastic moduli of the large rock image. There are two main advantages of applying the surrogate CNN model to homogenize subimages and using extra DNS to homogenize the coarse grained image. First, using smaller subimages for the surrogate CNN model overcomes the GPU memory issue faced during training. Second, a large training dataset can be constructed to train the surrogate CNN model. However, the size of the subimage should be large enough compared to the correlation length to capture the microstructure of the rock.

During the HHM, the majority of the computational time is spent during the homogenization of $(N/n)^3$ subimages. Replacing computationally expensive DNS with cheaper surrogate CNN model to homogenize the subimages gains a significant amount in computational time. For instance, six different simulations need to be carries out using independent strain boundary conditions to compute full $6\times 6$ effective stiffness matrix of a subimage of size $75\times 75 \times 75$ voxels which takes about 100 seconds. On the other hand, only one forward pass is necessary to compute effective elastic moduli using the trained surrogate CNN model, which takes about 0.1 second. Thus, using the CNN model represents a 1000 times gain in computational efficiency. Furthermore, the surrogate CNN model was trained using 2800 subimages and validated using 1000 images. On the other hand, the trained CNN model is used to homogenize about 20000 different subimages. Thus, the DNS need to be carried out only on one fifth of the total number of subimages to compose the training and validation datasets.

Another novelty of our proposed approach is that the surrogate CNN model make use of the Hashin-Shtrikman (HS) bounds to predict the effective bulk and shear moduli of the rock images. The final layer of the surrogate CNN model predicts two numbers each between 0 and 1 which represent the location of the effective moduli between the HS bounds. This feature of the CNN model increases the accuracy of the prediction and extends the transferability of the CNN model to rocks with different microstructure and mineral compositions.

The surrogate CNN model is first pretrained on a mixed dataset comprising images of five different sandstone rocks with quartz mineralogy. The performance of the surrogate CNN model pretrained on the mixed dataset is shown in \ref{app:pretrained_model}. Although the pretrained model predicts the effective elastic moduli of the sandstone rocks in decent agreement with the DNS for subimages of size $75\times 75\times 75$ voxels, the prediction of the pretrained CNN model becomes poor when applied to the images of size $N=300$ within CNN-HHM approach. To achieve better performance, we employ transfer learning to fine-tune the surrogate CNN model for each sandstone. The accuracy of the fine-tuned CNN model is shown to increase significantly when applied to the images of size $N=300$ and $N=600$ within CNN-HHM approach.

We also show that the proposed CNN-HHM approach$-$ with the surrogate CNN model fine-tuned using transfer learning $-$ can predict the effective elastic moduli of the rocks with different mineralogical compositions. We, specifically, fine-tune the surrogate CNN model for limestone rock which has calcite mineralogy and significantly different microstructure compared to the sandstone rocks. The fine-tuned model's prediction of effective elastic moduli of the limestone rocks is shown to be in good agreement with the DNS.

In this work, we employ 3D CNN model as a surrogate to predict homogenized elastic moduli of small subimages. CNN models can process images of fixed size. The variable size images can, however, be processes using graph neural networks (GNNs)~\citep{scarselliGraphNeuralNetwork2009,wuComprehensiveSurveyGraph2021} or vision transformer (ViT)~\citep{cordonnierRelationshipSelfAttentionConvolutional2020,dosovitskiyImageWorth16x162021} based surrogate models.

This work shows that an expensive DNS can be replaced with a cheaper surrogate CNN model within the framework of HHM without introducing any systematic error. The HHM, being an approximate method, itself introduces an error which, however, can be controlled and eliminated by using different size of subimages to perform the HHM~\citep{ahmadComputationEffectiveElastic2023}. Thus, multiple CNN models can be trained for multiple subimage sizes and utilize the procedure outlined in ~\citet{ahmadComputationEffectiveElastic2023} to reduce the HHM error.

Although this work focuses solely on predicting effective elastic moduli of large rock images, similar framework can be applied to homogenize other properties, such as permeability~\cite{Liu2022}, heat conductivity, electrical and magnetic properties. Thus, the proposed CNN-HHM approach can overcome the memory issues and enable an efficient estimation of effective properties from large rock images.

\appendix
\section{Performance of the pretrained surrogate CNN model}

In the main text, we show results of applying fine-tunes surrogate CNN model to homogenize elastic moduli of rock images. In this appendix, we present the results obtained by using surrogate CNN model pretrained on mixed sandstone dataset without any fine-tuning. Thus, all the results shown in this appendix are obtained from a single surrogate CNN model.

\label{app:pretrained_model}
\begin{figure}[ht!]
    \centering
    \includegraphics[width=\textwidth]{"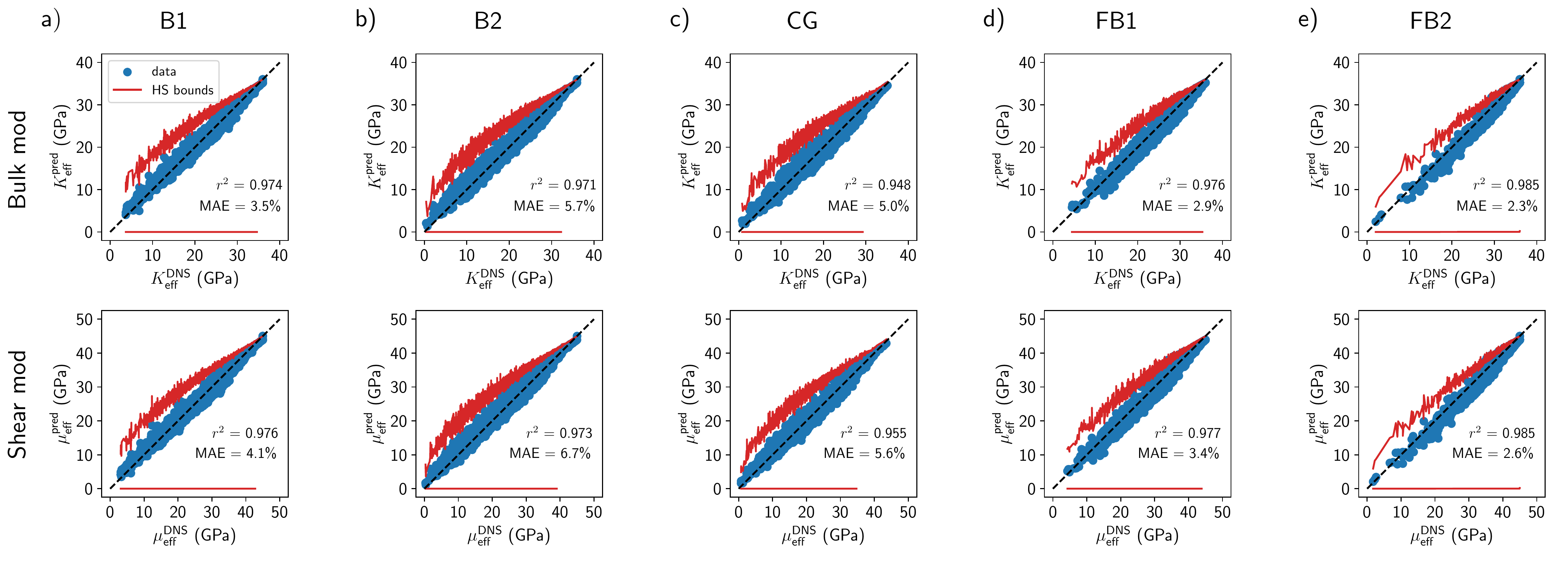"}
    \caption{Parity plot of the effective bulk and shear moduli predicted from the pretrained surrogate CNN model vs that obtained from DNS for 4096 subimages of size $75 \times 75 \times 75$ voxels. The results are shown for five sandstone rocks: (a) B1, (b) B2, (c) CG, (d) FB1, and (e) FB2. The blue dots represent the elastic moduli of the individual rocks and the black dashed line is $45^\circ$ line. Each figure, additionally, contains $r^2$ value and mean absolute error (MAE) of the prediction.}
    \label{fig:pretrain_75}
\end{figure}

Figure~\ref{fig:pretrain_75} compares the effective bulk and shear moduli obtained from the pretrained CNN model with that obtained from DNS for the five sandstone subimage of size $75\times 75\times 75$ voxels. The results show that a single pretrained CNN model is able to predict the effective elastic moduli of all five rocks in a reasonable agreement with the DNS. $r^2$ values remain above 0.948, and MAE lower than 6.7 \% across all five rocks. A comparison with fine-tuned CNN model, Figure~\ref{fig:test_result_fine_tune_True}, shows that the fine-tuning using transfer learning does not represent a significant improvement in the performance of the surrogate CNN model for the subimages of size $75\times 75\times 75$ voxels.

\begin{figure}[ht!]
    \centering
    \includegraphics[width=\textwidth]{"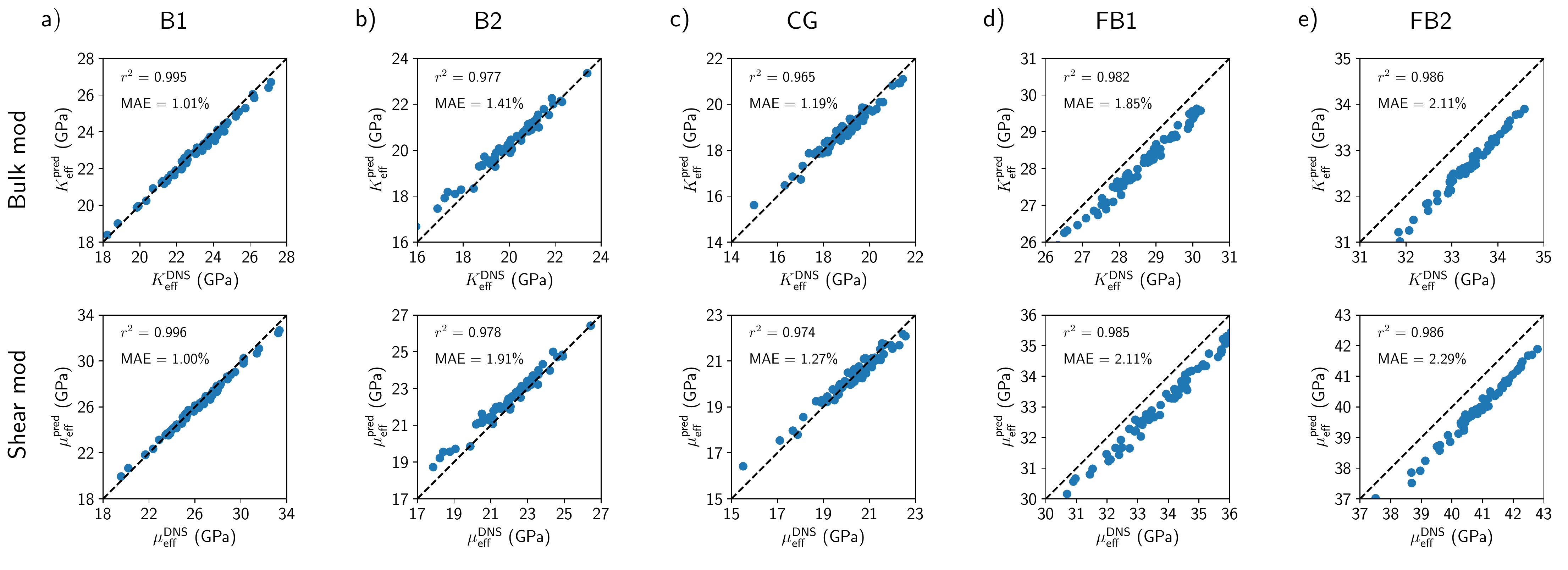"}
    \caption{Parity plot of the effective elastic moduli predicted from the CNN-HHM approach vs that obtained from DNS for the images of size $300\times 300\times 300$ voxels. The surrogate CNN model used here is not fine-tuned using transfer learning. The results are shown for five different sandstones: (a) B1, (b) B2, (c) CG, (d) FB1, and (e) FB2. The blue dots represent the elastic moduli of the individual rocks and the black dashed line is $45^\circ$ line. Each figure, additionally, contains $r^2$ value and mean absolute error (MAE) of the
        prediction.}
    \label{fig:pretrain_300}
\end{figure}

Figure~\ref{fig:pretrain_300} compares the effective bulk and shear moduli obtained from the pretrained CNN model with that obtained from DNS for the five sandstone images of size $300\times 300\times 300$ voxels. Although the pretrained CNN model is able to predict the effective elastic moduli of B1, B2, and CG rocks in a good agreement with the DNS, the pretrained CNN model performs poorly for FB1 and FB2 rocks. The pretrained CNN introduces a systematic error in the effective moduli of the two rocks.

\begin{figure}[ht!]
    \centering
    \includegraphics[width=0.6\textwidth]{"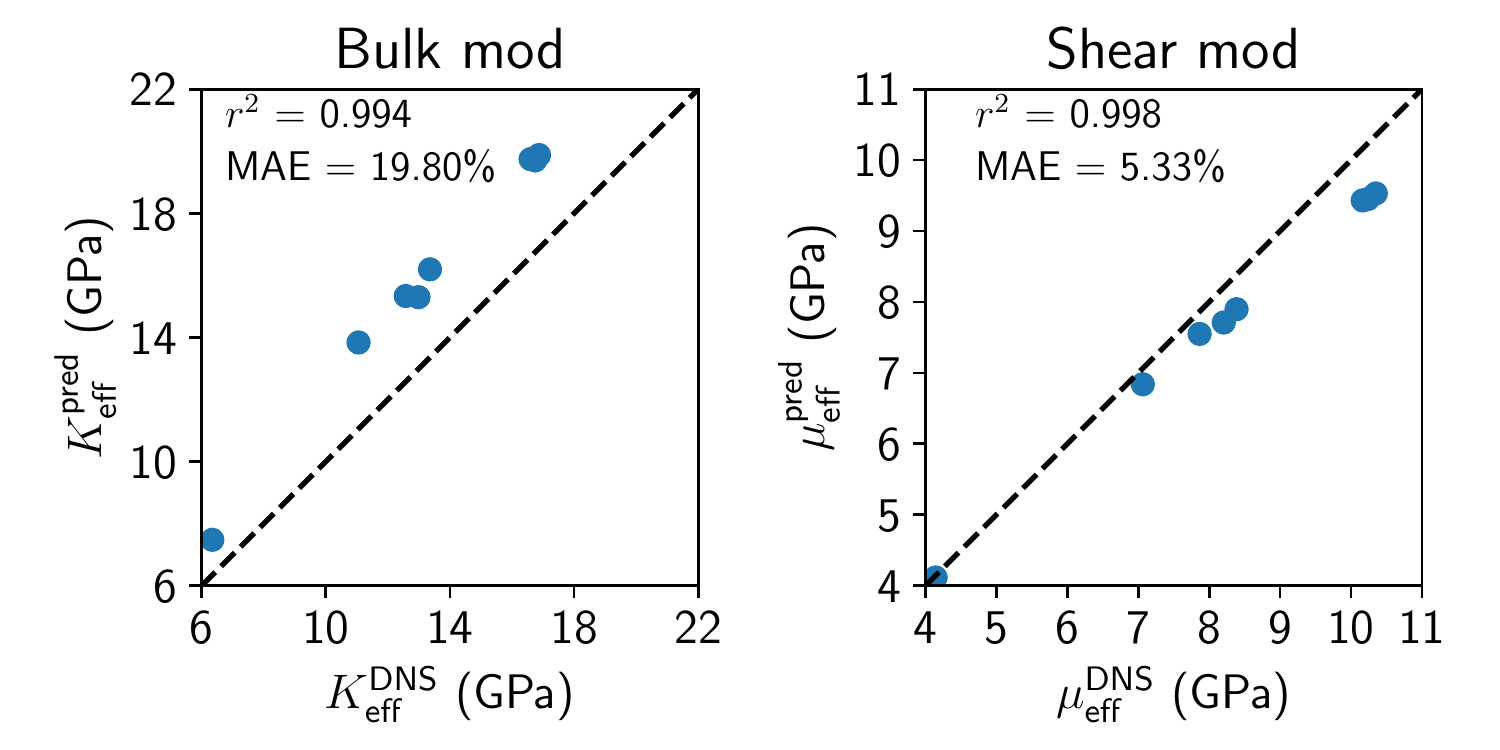"}
    \caption{Parity plot of the effective bulk and shear moduli predicted from the CNN-HHM approach vs that obtained from DNS for the 8 limestone rock images of size $300 \times 300 \times 300$ voxels. The surrogate CNN model used here is trained on sandstones, and is not been fine-tuned for limestone microstructure and mineralogy. The blue dots represent the elastic moduli of the individual rocks and the black dashed line is $45^\circ$ line. Each figure, additionally, contains $r^2$ value and mean absolute error (MAE) of the prediction.}
    \label{fig:pretrain_300_limestone}
\end{figure}

Finally, Figure~\ref{fig:pretrain_300_limestone} compares the effective bulk and shear moduli obtained from the pretrained CNN model with that obtained from DNS for the eight limestone rock images of size $300\times 300\times 300$ voxels.

Form the above results, it can be inferred that although a single pretrained surrogate CNN model is not able to predict the effective elastic moduli reasonably well, it seems to learn relevant features and capture a general mapping from microstructure to effective moduli. This is reflected from the fact that the pretrained predictions show a strong correlation with DNS values. Thus, optimizing only the fully connected layer during fine-tuning is sufficient to improve the performance of the surrogate CNN model for the specific rock type.

\addcontentsline{toc}{section}{References}
\bibliographystyle{model1-num-names}

\begin{thebibliography}{48}
    \expandafter\ifx\csname natexlab\endcsname\relax\def\natexlab#1{#1}\fi
    \providecommand{\bibinfo}[2]{#2}
    \ifx\xfnm\relax \def\xfnm[#1]{\unskip,\space#1}\fi
    \bibitem[{Arns et~al.(2001)Arns, Knackstedt, Pinczewski, and {W. B.
                        Lindquist}}]{Arns2001}
    \bibinfo{author}{C.~H. Arns}, \bibinfo{author}{M.~A. Knackstedt},
    \bibinfo{author}{W.~V. Pinczewski}, \bibinfo{author}{{W. B. Lindquist}},
    \newblock \bibinfo{title}{Accurate estimation of transport properties from
        microtomographic images},
    \newblock \bibinfo{journal}{Geophysics Research Letters} \bibinfo{volume}{28}
    (\bibinfo{year}{2001}) \bibinfo{pages}{3361--3364}.
    \bibitem[{Dvorkin et~al.(2011)Dvorkin, Derzhi, Diaz, and Fang}]{Dvorkin2011}
    \bibinfo{author}{J.~Dvorkin}, \bibinfo{author}{N.~Derzhi},
    \bibinfo{author}{E.~Diaz}, \bibinfo{author}{Q.~Fang},
    \newblock \bibinfo{title}{Relevance of computational rock physics},
    \newblock \bibinfo{journal}{Geophysics} \bibinfo{volume}{76}
    (\bibinfo{year}{2011}).
    \bibitem[{Andr{\"a} et~al.(2013{\natexlab{a}})Andr{\"a}, Combaret, Dvorkin,
                Glatt, Han, Kabel, Keehm, Krzikalla, Lee, Madonna, Marsh, Mukerji, Saenger,
                Sain, Saxena, Ricker, Wiegmann, and Zhan}]{Andra2013}
    \bibinfo{author}{H.~Andr{\"a}}, \bibinfo{author}{N.~Combaret},
    \bibinfo{author}{J.~Dvorkin}, \bibinfo{author}{E.~Glatt},
    \bibinfo{author}{J.~Han}, \bibinfo{author}{M.~Kabel},
    \bibinfo{author}{Y.~Keehm}, \bibinfo{author}{F.~Krzikalla},
    \bibinfo{author}{M.~Lee}, \bibinfo{author}{C.~Madonna},
    \bibinfo{author}{M.~Marsh}, \bibinfo{author}{T.~Mukerji},
    \bibinfo{author}{E.~H. Saenger}, \bibinfo{author}{R.~Sain},
    \bibinfo{author}{N.~Saxena}, \bibinfo{author}{S.~Ricker},
    \bibinfo{author}{A.~Wiegmann}, \bibinfo{author}{X.~Zhan},
    \newblock \bibinfo{title}{Digital rock physics benchmarks-{{Part I}}:
    {{Imaging}} and segmentation},
    \newblock \bibinfo{journal}{Computers and Geosciences} \bibinfo{volume}{50}
    (\bibinfo{year}{2013}{\natexlab{a}}) \bibinfo{pages}{25--32}.
    \bibitem[{Andr{\"a} et~al.(2013{\natexlab{b}})Andr{\"a}, Combaret, Dvorkin,
                Glatt, Han, Kabel, Keehm, Krzikalla, Lee, Madonna, Marsh, Mukerji, Saenger,
                Sain, Saxena, Ricker, Wiegmann, and Zhan}]{Andra2013a}
    \bibinfo{author}{H.~Andr{\"a}}, \bibinfo{author}{N.~Combaret},
    \bibinfo{author}{J.~Dvorkin}, \bibinfo{author}{E.~Glatt},
    \bibinfo{author}{J.~Han}, \bibinfo{author}{M.~Kabel},
    \bibinfo{author}{Y.~Keehm}, \bibinfo{author}{F.~Krzikalla},
    \bibinfo{author}{M.~Lee}, \bibinfo{author}{C.~Madonna},
    \bibinfo{author}{M.~Marsh}, \bibinfo{author}{T.~Mukerji},
    \bibinfo{author}{E.~H. Saenger}, \bibinfo{author}{R.~Sain},
    \bibinfo{author}{N.~Saxena}, \bibinfo{author}{S.~Ricker},
    \bibinfo{author}{A.~Wiegmann}, \bibinfo{author}{X.~Zhan},
    \newblock \bibinfo{title}{Digital rock physics benchmarks-part {{II}}:
    {{Computing}} effective properties},
    \newblock \bibinfo{journal}{Computers and Geosciences} \bibinfo{volume}{50}
    (\bibinfo{year}{2013}{\natexlab{b}}) \bibinfo{pages}{33--43}.
    \bibitem[{Saxena et~al.(2017)Saxena, Hofmann, Alpak, Dietderich, Hunter, and
                    {Day-Stirrat}}]{Saxena2017}
    \bibinfo{author}{N.~Saxena}, \bibinfo{author}{R.~Hofmann},
    \bibinfo{author}{F.~O. Alpak}, \bibinfo{author}{J.~Dietderich},
    \bibinfo{author}{S.~Hunter}, \bibinfo{author}{R.~J. {Day-Stirrat}},
    \newblock \bibinfo{title}{Effect of image segmentation \textbackslash\& voxel
    size on micro-{{CT}} computed effective transport \textbackslash\& elastic
    properties},
    \newblock \bibinfo{journal}{Marine and Petroleum Geology} \bibinfo{volume}{86}
    (\bibinfo{year}{2017}) \bibinfo{pages}{972--990}.
    \bibitem[{Saxena et~al.(2019)Saxena, Hofmann, Hows, Saenger, Duranti, Stefani,
                Wiegmann, Kerimov, and Kabel}]{Saxena2019}
    \bibinfo{author}{N.~Saxena}, \bibinfo{author}{R.~Hofmann},
    \bibinfo{author}{A.~Hows}, \bibinfo{author}{E.~H. Saenger},
    \bibinfo{author}{L.~Duranti}, \bibinfo{author}{J.~Stefani},
    \bibinfo{author}{A.~Wiegmann}, \bibinfo{author}{A.~Kerimov},
    \bibinfo{author}{M.~Kabel},
    \newblock \bibinfo{title}{Rock compressibility from microcomputed tomography
    images: {{Controls}} on digital rock simulations},
    \newblock \bibinfo{journal}{Geophysics} \bibinfo{volume}{84}
    (\bibinfo{year}{2019}) \bibinfo{pages}{WA127--WA139}.
    \bibitem[{Liu and Mukerji(2022)}]{Liu2022a}
    \bibinfo{author}{M.~Liu}, \bibinfo{author}{T.~Mukerji},
    \newblock \bibinfo{title}{Multiscale {{Fusion}} of {{Digital Rock Images
                        Based}} on {{Deep Generative Adversarial Networks}}},
    \newblock \bibinfo{journal}{Geophysical Research Letters} \bibinfo{volume}{49}
    (\bibinfo{year}{2022}).
    \bibitem[{Ahmad et~al.(2023)Ahmad, Liu, Ortiz, Mukerji, and
                Cai}]{ahmadComputationEffectiveElastic2023}
    \bibinfo{author}{R.~Ahmad}, \bibinfo{author}{M.~Liu},
    \bibinfo{author}{M.~Ortiz}, \bibinfo{author}{T.~Mukerji},
    \bibinfo{author}{W.~Cai},
    \newblock \bibinfo{title}{Computation of effective elastic moduli of rocks
        using hierarchical homogenization},
    \newblock \bibinfo{journal}{Journal of the Mechanics and Physics of Solids}
    \bibinfo{volume}{174} (\bibinfo{year}{2023}) \bibinfo{pages}{105268}.
    \bibitem[{Karimpouli and Tahmasebi(2019)}]{Karimpouli2019a}
    \bibinfo{author}{S.~Karimpouli}, \bibinfo{author}{P.~Tahmasebi},
    \newblock \bibinfo{title}{Segmentation of digital rock images using deep
        convolutional autoencoder networks},
    \newblock \bibinfo{journal}{Computers and Geosciences} \bibinfo{volume}{126}
    (\bibinfo{year}{2019}) \bibinfo{pages}{142--150}.
    \bibitem[{Phan et~al.(2021)Phan, Ruspini, and Lindseth}]{Phan2021}
    \bibinfo{author}{J.~Phan}, \bibinfo{author}{L.~C. Ruspini},
    \bibinfo{author}{F.~Lindseth},
    \newblock \bibinfo{title}{Automatic segmentation tool for {{3D}} digital rocks
        by deep learning},
    \newblock \bibinfo{journal}{Scientific Reports} \bibinfo{volume}{11}
    (\bibinfo{year}{2021}) \bibinfo{pages}{1--15}.
    \bibitem[{Cao et~al.(2022)Cao, Ji, Cui, and Liu}]{Cao2022}
    \bibinfo{author}{D.~Cao}, \bibinfo{author}{S.~Ji}, \bibinfo{author}{R.~Cui},
    \bibinfo{author}{Q.~Liu},
    \newblock \bibinfo{title}{Multi-task learning for digital rock segmentation and
        characteristic parameters computation},
    \newblock \bibinfo{journal}{Journal of Petroleum Science and Engineering}
    \bibinfo{volume}{208} (\bibinfo{year}{2022}) \bibinfo{pages}{109202}.
    \bibitem[{Niu et~al.(2020)Niu, Mostaghimi, Shabaninejad, Swietojanski, and
                Armstrong}]{Niu2020}
    \bibinfo{author}{Y.~Niu}, \bibinfo{author}{P.~Mostaghimi},
    \bibinfo{author}{M.~Shabaninejad}, \bibinfo{author}{P.~Swietojanski},
    \bibinfo{author}{R.~T. Armstrong},
    \newblock \bibinfo{title}{Digital {{Rock Segmentation}} for {{Petrophysical
                        Analysis With Reduced User Bias Using Convolutional Neural Networks}}},
    \newblock \bibinfo{journal}{Water Resources Research} \bibinfo{volume}{56}
    (\bibinfo{year}{2020}) \bibinfo{pages}{1--11}.
    \bibitem[{Wu et~al.(2018)Wu, Yin, and Xiao}]{Wu2018b}
    \bibinfo{author}{J.~Wu}, \bibinfo{author}{X.~Yin}, \bibinfo{author}{H.~Xiao},
    \newblock \bibinfo{title}{Seeing permeability from images: Fast prediction with
        convolutional neural networks},
    \newblock \bibinfo{journal}{Science Bulletin} \bibinfo{volume}{63}
    (\bibinfo{year}{2018}) \bibinfo{pages}{1215--1222}.
    \bibitem[{{Araya-Polo} et~al.(2020){Araya-Polo}, Alpak, Hunter, Hofmann, and
                Saxena}]{Araya-Polo2020}
    \bibinfo{author}{M.~{Araya-Polo}}, \bibinfo{author}{F.~O. Alpak},
    \bibinfo{author}{S.~Hunter}, \bibinfo{author}{R.~Hofmann},
    \bibinfo{author}{N.~Saxena},
    \newblock \bibinfo{title}{Deep learning\textendash driven permeability
        estimation from {{2D}} images},
    \newblock \bibinfo{journal}{Computational Geosciences} \bibinfo{volume}{24}
    (\bibinfo{year}{2020}) \bibinfo{pages}{571--580}.
    \bibitem[{Tian et~al.(2020)Tian, Qi, Sun, and Yaseen}]{Tian2020}
    \bibinfo{author}{J.~Tian}, \bibinfo{author}{C.~Qi}, \bibinfo{author}{Y.~Sun},
    \bibinfo{author}{Z.~M. Yaseen},
    \newblock \bibinfo{title}{Surrogate permeability modelling of low-permeable
        rocks using convolutional neural networks},
    \newblock \bibinfo{journal}{Computer Methods in Applied Mechanics and
        Engineering} \bibinfo{volume}{366} (\bibinfo{year}{2020})
    \bibinfo{pages}{113103}.
    \bibitem[{Santos et~al.(2020)Santos, Xu, Jo, Landry, Prodanovi{\'c}, and
                Pyrcz}]{Santos2020}
    \bibinfo{author}{J.~E. Santos}, \bibinfo{author}{D.~Xu},
    \bibinfo{author}{H.~Jo}, \bibinfo{author}{C.~J. Landry},
    \bibinfo{author}{M.~Prodanovi{\'c}}, \bibinfo{author}{M.~J. Pyrcz},
    \newblock \bibinfo{title}{{{PoreFlow-Net}}: {{A 3D}} convolutional neural
    network to predict fluid flow through porous media},
    \newblock \bibinfo{journal}{Advances in Water Resources} \bibinfo{volume}{138}
    (\bibinfo{year}{2020}).
    \bibitem[{Santos et~al.(2021)Santos, Yin, Jo, Pan, Kang, Viswanathan,
                Prodanovi{\'c}, Pyrcz, and Lubbers}]{Santos2021}
    \bibinfo{author}{J.~E. Santos}, \bibinfo{author}{Y.~Yin},
    \bibinfo{author}{H.~Jo}, \bibinfo{author}{W.~Pan}, \bibinfo{author}{Q.~Kang},
    \bibinfo{author}{H.~S. Viswanathan}, \bibinfo{author}{M.~Prodanovi{\'c}},
    \bibinfo{author}{M.~J. Pyrcz}, \bibinfo{author}{N.~Lubbers},
    \bibinfo{title}{Computationally {{Efficient Multiscale Neural Networks
                        Applied}} to {{Fluid Flow}} in {{Complex 3D Porous Media}}}, volume
    \bibinfo{volume}{140}, \bibinfo{publisher}{{Springer Netherlands}},
    \bibinfo{year}{2021}.
    \bibitem[{Rizk et~al.(2021)Rizk, Tembely, AlAmeri, and {Al-Shalabi}}]{Rizk2021}
    \bibinfo{author}{A.~S. Rizk}, \bibinfo{author}{M.~Tembely},
    \bibinfo{author}{W.~AlAmeri}, \bibinfo{author}{E.~W. {Al-Shalabi}},
    \newblock \bibinfo{title}{A {{Critical Literature Review}} on {{Rock
                        Petrophysical Properties Estimation}} from {{Images Based}} on {{Direct
                        Simulation}} and {{Machine Learning Techniques}}},
    \newblock \bibinfo{journal}{Society of Petroleum Engineers - Abu Dhabi
        International Petroleum Exhibition and Conference, ADIP 2021}
    (\bibinfo{year}{2021}).
    \bibitem[{Liu et~al.(2023)Liu, Ahmad, Cai, and
                Mukerji}]{liuHierarchicalHomogenizationDeepLearningBased2023}
    \bibinfo{author}{M.~Liu}, \bibinfo{author}{R.~Ahmad}, \bibinfo{author}{W.~Cai},
    \bibinfo{author}{T.~Mukerji},
    \newblock \bibinfo{title}{Hierarchical {{Homogenization With
                        Deep-Learning-Based Surrogate Model}} for {{Rapid Estimation}} of {{Effective
                        Permeability From Digital Rocks}}},
    \newblock \bibinfo{journal}{Journal of Geophysical Research: Solid Earth}
    \bibinfo{volume}{128} (\bibinfo{year}{2023}) \bibinfo{pages}{e2022JB025378}.
    \bibitem[{Cilli and Cilli(2022)}]{Cilli2022}
    \bibinfo{author}{P.~A. Cilli}, \bibinfo{author}{P.~A. Cilli},
    \newblock \bibinfo{title}{Machine learning for elastic-electrical
        cross-property modelling of sandstones}  (\bibinfo{year}{2022}).
    \bibitem[{Cui et~al.(2021)Cui, Cao, Liu, Zhu, and Jia}]{Cui2021}
    \bibinfo{author}{R.~Cui}, \bibinfo{author}{D.~Cao}, \bibinfo{author}{Q.~Liu},
    \bibinfo{author}{Z.~Zhu}, \bibinfo{author}{Y.~Jia},
    \newblock \bibinfo{title}{V {{Pand V Sprediction}} from digital rock images
        using a combination of {{U-Net}} and convolutional neural networks},
    \newblock \bibinfo{journal}{Geophysics} \bibinfo{volume}{86}
    (\bibinfo{year}{2021}) \bibinfo{pages}{MR27--MR37}.
    \bibitem[{Karimpouli and Tahmasebi(2019)}]{Karimpouli2019}
    \bibinfo{author}{S.~Karimpouli}, \bibinfo{author}{P.~Tahmasebi},
    \newblock \bibinfo{title}{Image-based velocity estimation of rock using
            {{Convolutional Neural Networks}}},
    \newblock \bibinfo{journal}{Neural Networks} \bibinfo{volume}{111}
    (\bibinfo{year}{2019}) \bibinfo{pages}{89--97}.
    \bibitem[{Saad et~al.(2019)Saad, Negara, and Ali}]{Saad2019}
    \bibinfo{author}{B.~Saad}, \bibinfo{author}{A.~Negara}, \bibinfo{author}{S.~S.
        Ali},
    \newblock \bibinfo{title}{Digital rock physics combined with machine learning
        for rock mechanical properties characterization},
    \newblock in: \bibinfo{booktitle}{Society of {{Petroleum Engineers}} - {{Abu
            Dhabi International Petroleum Exhibition}} and {{Conference}} 2018,
    {{ADIPEC}} 2018}.
    \bibitem[{Rabbani et~al.(2020)Rabbani, Babaei, Shams, Wang, and
                Chung}]{Rabbani2020}
    \bibinfo{author}{A.~Rabbani}, \bibinfo{author}{M.~Babaei},
    \bibinfo{author}{R.~Shams}, \bibinfo{author}{Y.~D. Wang},
    \bibinfo{author}{T.~Chung},
    \newblock \bibinfo{title}{{{DeePore}}: {{A}} deep learning workflow for rapid
    and comprehensive characterization of porous materials},
    \newblock \bibinfo{journal}{Advances in Water Resources} \bibinfo{volume}{146}
    (\bibinfo{year}{2020}) \bibinfo{pages}{103787}.
    \bibitem[{Eidel(2023)}]{Eidel2023}
    \bibinfo{author}{B.~Eidel},
    \newblock \bibinfo{title}{Deep {{CNNs}} as universal predictors of elasticity
        tensors in homogenization},
    \newblock \bibinfo{journal}{Computer Methods in Applied Mechanics and
        Engineering} \bibinfo{volume}{403} (\bibinfo{year}{2023})
    \bibinfo{pages}{115741}.
    \bibitem[{Wang et~al.(2019)Wang, Armstrong, and Mostaghimi}]{Wang2019a}
    \bibinfo{author}{Y.~D. Wang}, \bibinfo{author}{R.~T. Armstrong},
    \bibinfo{author}{P.~Mostaghimi},
    \newblock \bibinfo{title}{Enhancing {{Resolution}} of {{Digital Rock Images}}
        with {{Super Resolution Convolutional Neural Networks}}},
    \newblock \bibinfo{journal}{Journal of Petroleum Science and Engineering}
    \bibinfo{volume}{182} (\bibinfo{year}{2019}) \bibinfo{pages}{106261}.
    \bibitem[{Tawfik et~al.(2022)Tawfik, Adishesha, Hsi, Purswani, Johns, Shokouhi,
                Huang, and Karpyn}]{Tawfik2022}
    \bibinfo{author}{M.~S. Tawfik}, \bibinfo{author}{A.~S. Adishesha},
    \bibinfo{author}{Y.~Hsi}, \bibinfo{author}{P.~Purswani},
    \bibinfo{author}{R.~T. Johns}, \bibinfo{author}{P.~Shokouhi},
    \bibinfo{author}{X.~Huang}, \bibinfo{author}{Z.~T. Karpyn},
    \newblock \bibinfo{title}{Comparative {{Study}} of {{Traditional}} and
            {{Deep-Learning Denoising Approaches}} for {{Image-Based Petrophysical
                        Characterization}} of {{Porous Media}}},
    \newblock \bibinfo{journal}{Frontiers in Water} \bibinfo{volume}{3}
    (\bibinfo{year}{2022}).
    \bibitem[{Wang et~al.(2019)Wang, Teng, He, Feng, and Zhang}]{Wang2019b}
    \bibinfo{author}{Y.~Wang}, \bibinfo{author}{Q.~Teng}, \bibinfo{author}{X.~He},
    \bibinfo{author}{J.~Feng}, \bibinfo{author}{T.~Zhang},
    \newblock \bibinfo{title}{{{CT-image}} of rock samples super resolution using
            {{3D}} convolutional neural network},
    \newblock \bibinfo{journal}{Computers and Geosciences} \bibinfo{volume}{133}
    (\bibinfo{year}{2019}) \bibinfo{pages}{104314}.
    \bibitem[{Li et~al.(2018)Li, Zhang, Zhao, Burkhart, Brinson, and
                Chen}]{Li2018b}
    \bibinfo{author}{X.~Li}, \bibinfo{author}{Y.~Zhang}, \bibinfo{author}{H.~Zhao},
    \bibinfo{author}{C.~Burkhart}, \bibinfo{author}{L.~C. Brinson},
    \bibinfo{author}{W.~Chen},
    \newblock \bibinfo{title}{A {{Transfer Learning Approach}} for {{Microstructure
                        Reconstruction}} and {{Structure-property Predictions}}},
    \newblock \bibinfo{journal}{Scientific Reports} \bibinfo{volume}{8}
    (\bibinfo{year}{2018}) \bibinfo{pages}{1--13}.
    \bibitem[{Mosser et~al.(2018)Mosser, Dubrule, and Blunt}]{Mosser2018}
    \bibinfo{author}{L.~Mosser}, \bibinfo{author}{O.~Dubrule},
    \bibinfo{author}{M.~J. Blunt},
    \newblock \bibinfo{title}{Stochastic {{Reconstruction}} of an {{Oolitic
                        Limestone}} by {{Generative Adversarial Networks}}},
    \newblock \bibinfo{journal}{Transport in Porous Media} \bibinfo{volume}{125}
    (\bibinfo{year}{2018}) \bibinfo{pages}{81--103}.
    \bibitem[{{dos Anjos} et~al.(2021){dos Anjos}, Avila, Vasconcelos,
                Pereira~Neta, Medeiros, Evsukoff, Surmas, and Landau}]{DosAnjos2021}
    \bibinfo{author}{C.~E. {dos Anjos}}, \bibinfo{author}{M.~R. Avila},
    \bibinfo{author}{A.~G. Vasconcelos}, \bibinfo{author}{A.~M. Pereira~Neta},
    \bibinfo{author}{L.~C. Medeiros}, \bibinfo{author}{A.~G. Evsukoff},
    \bibinfo{author}{R.~Surmas}, \bibinfo{author}{L.~Landau},
    \newblock \bibinfo{title}{Deep learning for lithological classification of
    carbonate rock micro-{{CT}} images},
    \newblock \bibinfo{journal}{Computational Geosciences} \bibinfo{volume}{25}
    (\bibinfo{year}{2021}) \bibinfo{pages}{971--983}.
    \bibitem[{Kashefi and Mukerji(2021)}]{Kashefi2021}
    \bibinfo{author}{A.~Kashefi}, \bibinfo{author}{T.~Mukerji},
    \newblock \bibinfo{title}{Point-cloud deep learning of porous media for
        permeability prediction},
    \newblock \bibinfo{journal}{Physics of Fluids} \bibinfo{volume}{33}
    (\bibinfo{year}{2021}).
    \bibitem[{Hashin and Shtrikman(1963)}]{Hashin1963}
    \bibinfo{author}{Z.~Hashin}, \bibinfo{author}{S.~Shtrikman},
    \newblock \bibinfo{title}{A variational approach to the theory of the elastic
        behaviour of multiphase materials},
    \newblock \bibinfo{journal}{Journal of the Mechanics and Physics of Solids}
    \bibinfo{volume}{11} (\bibinfo{year}{1963}) \bibinfo{pages}{127--140}.
    \bibitem[{Otsu(1979)}]{otsuThresholdSelectionMethod1979}
    \bibinfo{author}{N.~Otsu},
    \newblock \bibinfo{title}{A {{Threshold Selection Method}} from {{Gray-Level
                        Histograms}}},
    \newblock \bibinfo{journal}{IEEE Transactions on Systems, Man, and Cybernetics}
    \bibinfo{volume}{9} (\bibinfo{year}{1979}) \bibinfo{pages}{62--66}.
    \bibitem[{Mavko et~al.(2003)Mavko, Mukerji, and Dvorkin}]{mavko2003rock}
    \bibinfo{author}{G.~Mavko}, \bibinfo{author}{T.~Mukerji},
    \bibinfo{author}{J.~Dvorkin}, \bibinfo{title}{The {{Rock Physics Handbook}}:
    {{Tools}} for {{Seismic Analysis}} of {{Porous Media}}},
    \bibinfo{publisher}{{Cambridge University Press}}, \bibinfo{year}{2003}.
    \bibitem[{Ela(????)}]{Elastodict}
    \bibinfo{title}{{{http://www.geodict.de/Modules/Dicts/ElastoDict.php.}}}, ????
    \bibitem[{Kabel and Andr{\"a}(2013)}]{Kabel2013}
    \bibinfo{author}{M.~Kabel}, \bibinfo{author}{H.~Andr{\"a}},
    \newblock \bibinfo{title}{Fast numerical computation of effective elastic
        moduli of porous materials},
    \newblock \bibinfo{journal}{Rep. Fraunhofer ITWM} \bibinfo{volume}{224}
    (\bibinfo{year}{2013}) \bibinfo{pages}{1--16}.
    \bibitem[{Kabel et~al.(2016)Kabel, Fliegener, and Schneider}]{Kabel2016}
    \bibinfo{author}{M.~Kabel}, \bibinfo{author}{S.~Fliegener},
    \bibinfo{author}{M.~Schneider},
    \newblock \bibinfo{title}{Mixed boundary conditions for {{FFT-based}}
        homogenization at finite strains},
    \newblock \bibinfo{journal}{Computational Mechanics} \bibinfo{volume}{57}
    (\bibinfo{year}{2016}) \bibinfo{pages}{193--210}.
    \bibitem[{Schneider et~al.(2016)Schneider, Ospald, and Kabel}]{Schneider2016}
    \bibinfo{author}{M.~Schneider}, \bibinfo{author}{F.~Ospald},
    \bibinfo{author}{M.~Kabel},
    \newblock \bibinfo{title}{Computational homogenization of elasticity on a
        staggered grid},
    \newblock \bibinfo{journal}{International Journal for Numerical Methods in
        Engineering} \bibinfo{volume}{105} (\bibinfo{year}{2016})
    \bibinfo{pages}{693--720}.
    \bibitem[{Kingma and Ba(2017)}]{kingmaAdamMethodStochastic2017}
    \bibinfo{author}{D.~P. Kingma}, \bibinfo{author}{J.~Ba}, \bibinfo{title}{Adam:
    {{A Method}} for {{Stochastic Optimization}}}, \bibinfo{year}{2017}.
    \bibitem[{Paszke et~al.(2019)Paszke, Gross, Massa, Lerer, Bradbury, Chanan,
                Killeen, Lin, Gimelshein, Antiga, Desmaison, Kopf, Yang, DeVito, Raison,
                Tejani, Chilamkurthy, Steiner, Fang, Bai, and Chintala}]{NIPS2019_9015}
    \bibinfo{author}{A.~Paszke}, \bibinfo{author}{S.~Gross},
    \bibinfo{author}{F.~Massa}, \bibinfo{author}{A.~Lerer},
    \bibinfo{author}{J.~Bradbury}, \bibinfo{author}{G.~Chanan},
    \bibinfo{author}{T.~Killeen}, \bibinfo{author}{Z.~Lin},
    \bibinfo{author}{N.~Gimelshein}, \bibinfo{author}{L.~Antiga},
    \bibinfo{author}{A.~Desmaison}, \bibinfo{author}{A.~Kopf},
    \bibinfo{author}{E.~Yang}, \bibinfo{author}{Z.~DeVito},
    \bibinfo{author}{M.~Raison}, \bibinfo{author}{A.~Tejani},
    \bibinfo{author}{S.~Chilamkurthy}, \bibinfo{author}{B.~Steiner},
    \bibinfo{author}{L.~Fang}, \bibinfo{author}{J.~Bai},
    \bibinfo{author}{S.~Chintala},
    \newblock \bibinfo{title}{{{PyTorch}}: {{An Imperative Style}},
    {{High-Performance Deep Learning Library}}},
    \newblock in: \bibinfo{editor}{H.~Wallach}, \bibinfo{editor}{H.~Larochelle},
    \bibinfo{editor}{A.~Beygelzimer}, \bibinfo{editor}{F.~{D'Alch{\'e}-Buc}},
    \bibinfo{editor}{E.~Fox}, \bibinfo{editor}{R.~Garnett} (Eds.),
    \bibinfo{booktitle}{Advances in {{Neural Information Processing Systems}}
        32}, \bibinfo{publisher}{{Curran Associates, Inc.}}, \bibinfo{year}{2019},
    pp. \bibinfo{pages}{8024--8035}.
    \bibitem[{Moulinec and Suquet(1994)}]{Moulinec1994}
    \bibinfo{author}{H.~Moulinec}, \bibinfo{author}{P.~Suquet},
    \newblock \bibinfo{title}{A fast numerical method for computing the linear and
        nonlinear mechanical properties of composites},
    \newblock \bibinfo{journal}{Mechanics of solids}  (\bibinfo{year}{1994}).
    \bibitem[{{de Geus} et~al.(2017){de Geus}, Vond{\v r}ejc, Zeman, Peerlings, and
                Geers}]{DeGeus2017a}
    \bibinfo{author}{T.~W. {de Geus}}, \bibinfo{author}{J.~Vond{\v r}ejc},
    \bibinfo{author}{J.~Zeman}, \bibinfo{author}{R.~H. Peerlings},
    \bibinfo{author}{M.~G. Geers},
    \newblock \bibinfo{title}{Finite strain {{FFT-based}} non-linear solvers made
        simple},
    \newblock \bibinfo{journal}{Computer Methods in Applied Mechanics and
        Engineering} \bibinfo{volume}{318} (\bibinfo{year}{2017})
    \bibinfo{pages}{412--430}.
    \bibitem[{Scarselli et~al.(2009)Scarselli, Gori, Tsoi, Hagenbuchner, and
                Monfardini}]{scarselliGraphNeuralNetwork2009}
    \bibinfo{author}{F.~Scarselli}, \bibinfo{author}{M.~Gori},
    \bibinfo{author}{A.~C. Tsoi}, \bibinfo{author}{M.~Hagenbuchner},
    \bibinfo{author}{G.~Monfardini},
    \newblock \bibinfo{title}{The {{Graph Neural Network Model}}},
    \newblock \bibinfo{journal}{IEEE Transactions on Neural Networks}
    \bibinfo{volume}{20} (\bibinfo{year}{2009}) \bibinfo{pages}{61--80}.
    \bibitem[{Wu et~al.(2021)Wu, Pan, Chen, Long, Zhang, and
                Yu}]{wuComprehensiveSurveyGraph2021}
    \bibinfo{author}{Z.~Wu}, \bibinfo{author}{S.~Pan}, \bibinfo{author}{F.~Chen},
    \bibinfo{author}{G.~Long}, \bibinfo{author}{C.~Zhang}, \bibinfo{author}{P.~S.
        Yu},
    \newblock \bibinfo{title}{A {{Comprehensive Survey}} on {{Graph Neural
                        Networks}}},
    \newblock \bibinfo{journal}{IEEE Trans. Neural Netw. Learning Syst.}
    \bibinfo{volume}{32} (\bibinfo{year}{2021}) \bibinfo{pages}{4--24}.
    \bibitem[{Cordonnier et~al.(2020)Cordonnier, Loukas, and
                Jaggi}]{cordonnierRelationshipSelfAttentionConvolutional2020}
    \bibinfo{author}{J.-B. Cordonnier}, \bibinfo{author}{A.~Loukas},
    \bibinfo{author}{M.~Jaggi}, \bibinfo{title}{On the {{Relationship}} between
            {{Self-Attention}} and {{Convolutional Layers}}}, \bibinfo{year}{2020}.
    \bibitem[{Dosovitskiy et~al.(2021)Dosovitskiy, Beyer, Kolesnikov, Weissenborn,
                Zhai, Unterthiner, Dehghani, Minderer, Heigold, Gelly, Uszkoreit, and
                Houlsby}]{dosovitskiyImageWorth16x162021}
    \bibinfo{author}{A.~Dosovitskiy}, \bibinfo{author}{L.~Beyer},
    \bibinfo{author}{A.~Kolesnikov}, \bibinfo{author}{D.~Weissenborn},
    \bibinfo{author}{X.~Zhai}, \bibinfo{author}{T.~Unterthiner},
    \bibinfo{author}{M.~Dehghani}, \bibinfo{author}{M.~Minderer},
    \bibinfo{author}{G.~Heigold}, \bibinfo{author}{S.~Gelly},
    \bibinfo{author}{J.~Uszkoreit}, \bibinfo{author}{N.~Houlsby},
    \bibinfo{title}{An {{Image}} is {{Worth}} 16x16 {{Words}}: {{Transformers}}
    for {{Image Recognition}} at {{Scale}}}, \bibinfo{year}{2021}.
    \bibitem[{Liu et~al.(2022)Liu, Ahmad, Cai, and Mukerji}]{Liu2022}
    \bibinfo{author}{M.~Liu}, \bibinfo{author}{R.~Ahmad}, \bibinfo{author}{W.~Cai},
    \bibinfo{author}{T.~Mukerji},
    \newblock \bibinfo{title}{Hierarchical homogenization with deep-learning-based
        surrogate model for rapid estimation of effective permeability from digital
        rocks},
    \newblock \bibinfo{journal}{Earth and Space Science Open Archive}
    (\bibinfo{year}{2022}).

\end{thebibliography}

\end{document}